%% file: main.tex
\documentclass[conference]{IEEEtran}
\IEEEoverridecommandlockouts
\usepackage{cite}
\usepackage{amsmath,amssymb,amsfonts}
\usepackage{algorithm2e}
\usepackage{algpseudocode}
\usepackage{graphicx}
\usepackage{textcomp}
\usepackage{xcolor}
\usepackage{defines}
\usepackage{comment}
\usepackage{subcaption}
\usepackage{url}
\usepackage{flushend}

\def\BibTeX{{\rm B\kern-.05em{\sc i\kern-.025em b}\kern-.08em
    T\kern-.1667em\lower.7ex\hbox{E}\kern-.125emX}}
\begin{document}

\title{
\name{}: Genomics Acceleration Beyond On-Device Memory
}

\author{
\IEEEauthorblockN{Se-Min Lim}
\IEEEauthorblockA{\textit{Department of Computer Science} \\
\textit{University of California, Irvine}\\
Irvine, CA, USA \\
seminl1@ics.uci.edu}
\and
\IEEEauthorblockN{Seongyoung Kang}
\IEEEauthorblockA{\textit{Department of Computer Science} \\
\textit{University of California, Irvine}\\
Irvine, CA, USA \\
seongyk3@uci.edu}
\and
\IEEEauthorblockN{Sang-Woo Jun}
\IEEEauthorblockA{\textit{Department of Computer Science} \\
\textit{University of California, Irvine}\\
Irvine, CA, USA \\
swjun@ics.uci.edu}
}

\maketitle

\input{abstract}


\input{introduction}

\input{background}

\input{compression}

\input{software}

\input{hardware}

\input{evaluation}

\section{Conclusion and Discussion}
\label{sec:conclusion}

In this paper, we present \name{}, a computational genomics accelerator that provides access to TBs of genomics data at memory-class performance via hardware-optimized compression.
Our prototype implementation overcomes on-device memory capacity while supporting 30+\% peak accelerator utilization, a stark improvement from the conventional 3 $\sim$ 4\% over uncompressed PCIe.
\name{} supports memory-class performance to practically infinite data capacity, while using a fixed amount of HBM.
We believe this is an attractive solution to achieving scalability with computational genomics.

\section*{Acknowledgements}
This research is partially supported by the PRISM (000705769) center
under the JUMP 2.0 program by DARPA/SRC.

\bibliographystyle{ieeetr}
\bibliography{references,references_genomics}

\end{document}

%% file: abstract.tex
\begin{abstract}

This paper presents \name{}, a computational genomics acceleration platform that provides the illusion of practically infinite on-device memory capacity by compressing genomic data movement over PCIe.
\name{} introduces novel optimizations for efficient accelerator implementation to reference-based genome compression, including fixed-stride matching using cuckoo hashes and grouped header encoding, incorporated into a familiar interface supporting random accesses.
We evaluate a prototype implementation of \name{} on an affordable Alveo U50 FPGA equipped with 8~GB of HBM.
Thanks to the orders of magnitude improvements in performance and resource efficiency of genomic compression, our prototype provides access to TBs of host-side genomic data at memory-class performance, measuring speeds over 30\% of the on-device HBM bandwidth, an order of magnitude higher than conventional PCIe-limited architectures.
Using a real-world pre-alignment filtering application, \name{} demonstrates over 6$\times$ improvement over the conventional PCIe-attached architecture, achieving 30\% of peak internal throughput of an accelerator with HBM, and 90\% of the one with DDR4.
\name{} supports memory-class performance to practically infinite data capacity, using a small, fixed amount of HBM, making it an attractive solution to continued future scalability of computational genomics.

\end{abstract}

%% file: introduction.tex
\section{Introduction}
\label{sec:introduction}

\begin{figure}[htb]
    \centering
    \includegraphics[width=\columnwidth,page=8]{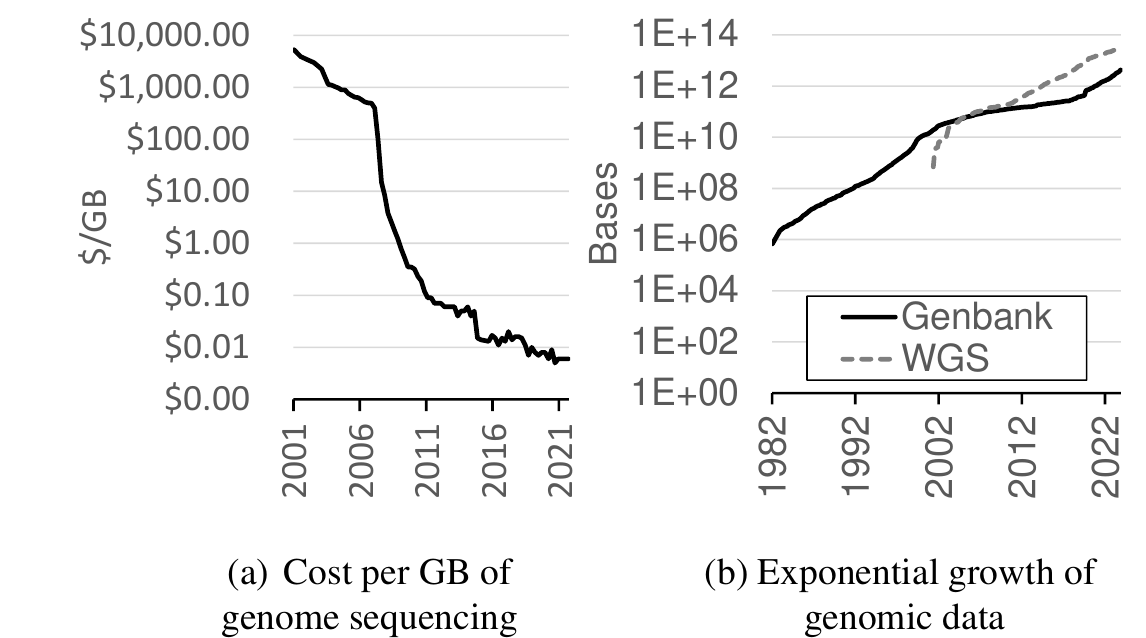}
    \caption{Exponentially cheaper sequencing leads to exponential growth in data.}
    \label{fig:growth}
\end{figure}

The rapid advancement of genome sequencing technologies has led to an explosion of collected genomic data, projected to become one of the largest and computationally demanding areas of computing in the next decade~\cite{stephens2015bigdatagenomics}.
Figure~\ref{fig:growth} presents both the exponential reduction of genome sequencing costs~\cite{dna_sequencing_cost}, and the corresponding exponentially increasing size of genomic databases~\cite{genbank_bases}.
In fact, the projected growth of genomic data is expected to outpace even Moore's law, growing faster than the rates of both computational power and memory capacity scaling~\cite{chen2020parc,kaplan2020bioseal,turakhia2018darwin}.

It is crucial to address the gap between data growth and computational capacity scaling, considering that some useful genomic analytics tasks are already too computationally expensive for real-world deployment.
Exceptional advancements beyond Moore's law must be made to ensure we can actually take advantage of the increasing availability of personal genomic data and rapid algorithmic advancements.
One representative example is De Novo sequence assembly, assembling genome reads among themselves without depending on a reference.
De Novo assembly can improve the quality of personalized genomic data analysis to levels sufficient for effective personalized medicine~\cite{ashley2016towardsprecisionmedicine,chaisson2015geneticdenovo}, but their computational overhead and cost are already considered too high to be viable in a clinical setting~\cite{ashley2016towardsprecisionmedicine}.

Because data growth is outpacing Moore's law, existing hardware and software solutions will not be sufficient to enable the benefits that genomics data may provide us.

One promising solution is computation accelerators as coprocessors.
A wide range of accelerators for genomics are being explored for various genomics applications, taking advantage of highly parallel fabrics including such as Graphic Processing Units (GPU)~\cite{guo2013gpu,liu2009cudasw++,liu2010cudasw++,jia2011metabing,liu2013cudasw++,cho2013xsd,su2014gpu,liu2015gswabe,wilton2015arioc,de2016cudalign,nishimura2017accelerating,kobus2017accelerating,houtgast2017efficient,cheng2018bitmapper2,ahmed2019gasal2,zeni2020logan,gamaarachchi2020gpu,kobus2021metacache,wang2023gpmeta,ju2024seedhitgpu}, Field-Programmable Gate Arrays (FPGA)~\cite{arram2015fpgareferencecompressiongenomic,chen2023efficientsequencingcompressionfpga,chen2014accelerating,zhao2017streaming,liao2018adaptively,qiao2019fpga,chen2021high,haghi2021fpga,li2021pipebsw,liyanage2023efficient}, Application-Specific Integrated Circuits (ASIC)~\cite{madhavan2014race,turakhia2017darwin,turakhia2018darwin,huangfu2018radar,fujiki2018genax,turakhia2019darwin}, Processing-In-Memory (PIM)~\cite{zokaee2018aligner,angizi2019aligns,gupta2019rapid,kaplan2020bioseal,chen2020parc,cali2020genasm,khatamifard2021genvom,li2021pim,wu2021sieve,shahroodi2022krakenonmem,shahroodi2022demeter,hanhan2022edam,zou2022biohd,jahshan2023dash,zhang2023aligner} and more~\cite{sarkar2019algorithm,sarkar2021quaser,mansouri2022genstore,varsamis2023quantum,wu2024abakus,ghiasi2024megis}.
These accelerators not only supply high computational throughput, but are also equipped with high-performance memory fabric, such as High-Bandwidth Memory (HBM), to satisfy both the computational and data access performance requirements of genomics applications.

Unfortunately, the performance of conventional accelerators equipped with high-performance on-device memory is often limited by the data movement overhead when the dataset exceeds on-device memory~\cite{zhang2019flashgpu,dhar2019near,lee2020optimizing,jonatan2024scalability}.
This is concerning because memory capacity scalability is unclear in the near future~\cite{kim2015architectural,shiratake2020scaling,kim2024present,hyun2024pathfinding}.
For workloads with low locality, every new access exceeding local memory must be transported over either PCIe or network links, resulting in orders of magnitude performance degradation due to orders of magnitude slower than 3D-stacked HBM.
Many important genomics applications fall into this category, including long read sequence alignment, pre-alignment filtering, overlap graph construction, and more.
In these applications, each read is loaded into accelerator memory, processed once, and then discarded to make space for the next data.
Indeed, many accelerator designs have pointed to data movement overhead as the primary bottleneck of performance~\cite{mailthody2019deepstore,liang2019ins,zhang2020dram,fang2020memory,preethi2020fpga,kim2021behemoth,singh2021fpga,zhou2022survey,knoben2023improving}.

\begin{figure}[htb]
    \centering
    \includegraphics[width=\columnwidth,page=2]{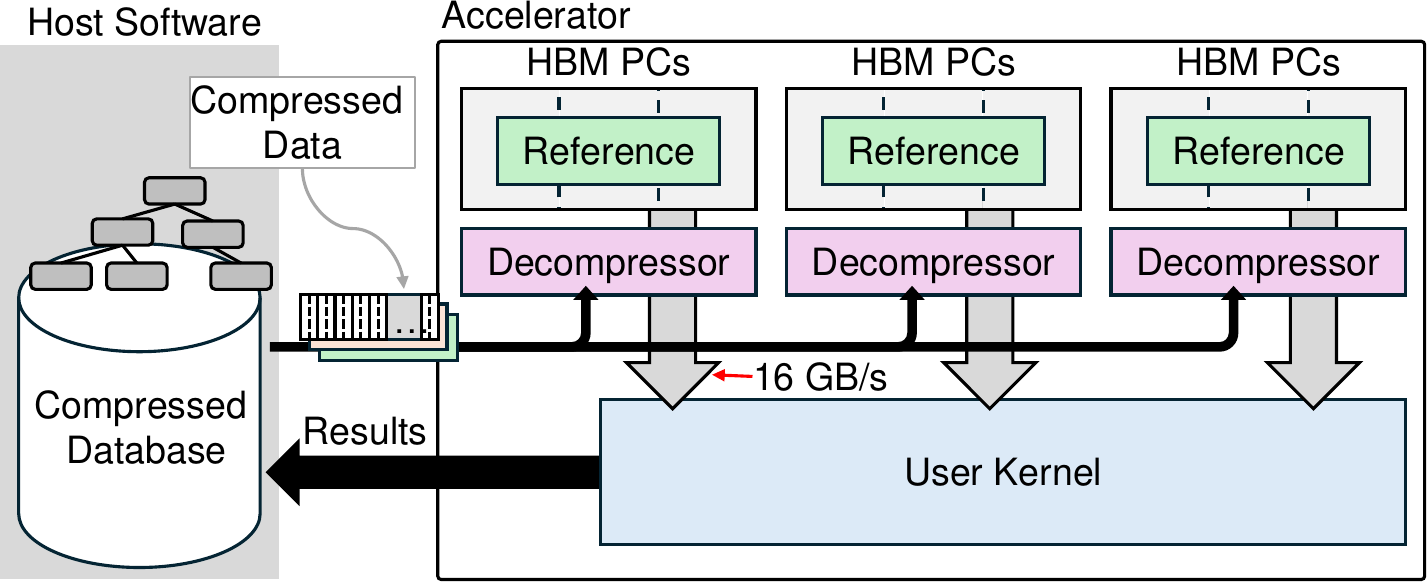}
    \caption{User kernels have access to unlimited high-performance memory capacity over compressed PCIe.}
    \label{fig:overall}
\end{figure}

In this work, we address these issues with \textbf{\name{}}, a genomics accelerator platform which provides application-specific hardware accelerators with high-bandwidth access to genomics data beyond on-device memory capacity.
\name{} introduces a novel, hardware-codesigned compression algorithm with low hardware overhead, whose hardware implementation works together with a software interface to mitigate the bandwidth gap between on-device memory and host-side PCIe or CXL/network interfaces.
As a result, \name{} provides practically unlimited capacity via NVMe storage or disaggregated memory, at DRAM-level bandwidth.
Figure~\ref{fig:overall} illustrates an example accelerator configuration which provides the user kernel with three parallel for data input.
The number of compressor/decompressor modules can be configured according to the requirements of the application.

The compression/decompression module of \name{} achieves multiple conflicting criteria, including compression efficiency, compression, and decompression throughput, and low hardware resource utilization, by using a reference-based algorithm with hardware-aware optimizations such as fixed-width grouped header encoding, fixed-k k-mer based matching as well as wide mismatch strides.
The compression module also uses a hardware-accelerated probabilistic filter to reduce the pressure on the memory system.
All of the hardware accelerators are managed by a host-side software management layer, which maintains the data structures for the random accesses into the genome data, as well as orchestrate the data movement between the compressed genome database and the user-provided application-specific hardware accelerator kernel.

We evaluate \name{} with an FPGA prototype using the Xilinx Alveo U50 board, on various genomic workloads, including pre-alignment filtering, using long read and reference genomes from various organisms.
The U50 is a useful and representative evaluation platform thanks to its realistic PCIe Gen~3 $\times$16 interface and 3D-stacked HBM memory, both of which are the typical limiting factors in data-intensive analytics.
It is also useful as a deployment target for \name{} by itself, since it is a mid-range FPGA affordable to many potential users, and FPGAs have often demonstrated superior performance and power efficiency compared to other programmable accelerators such as GPUs, on many representative genomics-related workloads~\cite{chen2016spark,fei2018fpgasw,rucci2018swifold,wu2019fpga,fujiki2020seedex}.

\name{} on the FPGA prototype can provide a user kernel with practically unlimited memory capacity, divided into multiple channels similar to HBM.
The physical constraints of the U50 FPGA allow up to ten channels, each delivering up to 16~GB/s of genomic data bandwidth, adding up to 160~GB/s bandwidth without running into the PCIe bottleneck.
This is about 31\% of the peak HBM performance, while consuming only 30\% of the chip space.
This is made possible by the hardware co-optimized compression algorithm achieving high compression ratios between 24$\times$ to 100+$\times$, competitive with state-of-the-art schemes.
We note that the power constraints specific to the U50 platform limit the accessible HBM bandwidth to less than half the peak bandwidth ($\sim$200~GB/s), also limiting the nominal \name{} bandwidth to 70~GB/s.
The throttled speed is still an order of magnitude higher than the PCIe bandwidth limitation, and near peak memory bandwidth that the DDR4 memory system can deliver on the higher-end U250 FPGA.
We note that this power throttling does not exist on other HBM-enabled FPGAs, such as the U55C or U280.
\name{} also achieves high compression performance of up to 3.7~GB/s, an order of magnitude faster than any existing FPGA genomic compression accelerator.

We also demonstrate overall system performance improvements on an important real-world workload of pre-alignment filtering for long-read alignment, and demonstrate almost 6$\times$ performance improvement over a conventional FPGA accelerator within the constraints of the U50, and almost 15$\times$ if we assume no HBM power throttling.

\textbf{Implications of \name{} are profound.}
Evaluation results show that using a small, fixed amount of on-device HBM memory, \name{} can deliver user accelerator kernels with memory-class bandwidth to practically unlimited capacity of genomics data via host-side NVMe, CXL, or disaggregated memory.
This has the attractive implication that achieving data scalability can be separated from memory capacity scalability, which is expected to be unfortunately difficult in the near future~\cite{kim2015architectural,shiratake2020scaling,kim2024present,hyun2024pathfinding}.

The rest of this paper is organized as follows:
First, we present background and related works in Section~\ref{sec:background}.
We present the \name{} compression and decompression algorithms in detail in Section~\ref{sec:algorithm}.
The software and hardware components of the \name{} system are described in Section~\ref{sec:software} and Section~\ref{sec:hardware}, respectively.
We provide evaluations and comparative results in Section~\ref{sec:evaluation}, and conclude with discussions in Section~\ref{sec:conclusion}.

%% file: background.tex
\section{Background and Related Works}
\label{sec:background}

\subsection{Computational Genomics and its Applications}

The increasingly wide availability of personal genomic data, coupled with rapid algorithmic advancements related to computational genomics, are enabling many important and useful applications, including personalized and precision medicine~\cite{nakagawa2018whole}, wide-scale genome-wide association studies
~\cite{gage2016gwascantellus,cano2020gwastofunction} for disease analysis, future agriculture~\cite{varshney2021designingfuturecrops}, and tracing and managing pandemics~\cite{sawyer2021metagenomics,racaniello2016movingbeyondmeta}.
Emerging solutions suggest highly effective personalized therapies for previously untreatable diseases may become available in the near future~\cite{ginsburg2009genomicpersonalized,offit2011personalizedmedicine}.

The fundamental task of computational genomics is to extract insight from the output available from modern sequencing technologies.
Because we cannot simply read genomic data from cell samples in a single complete string yet, what sequencing machines do is generate vast amounts of randomly sampled substrings from the whole genome.
These substrings are called ``reads'', and we cannot know where they were sampled from, and they may contain noise and other probabilistic errors.
Based on the average length of reads, sequencing technologies are categorized into short reads (hundreds or less) or long reads (thousands to millions).
Furthermore, since the DNA is a double helix consisting of two complementary strands, a read may come from either strand, resulting in either a ``forward'' read, or a ``reverse complement'' read.
A forward read can be computed from the reverse read and vice versa, and the existence of opposite directions in the sample must be considered during analysis.

The first task of computational genomics is typically to reconstruct the reads into a whole genome while removing noise and errors, typically using numerous overlapping samples, in a process called ``sequence assembly''.
Assembly can be done using a pre-assembled reference genome of an organism to take advantage of similarities between individuals (``reference-based'' assembly) or just between the reads themselves ``De Novo'' assembly), either because a reference is not available, or to avoid errors such as ``reference drift''~\cite{ashley2016towardsprecisionmedicine,chaisson2015geneticdenovo}.
Both approaches depend on ``sequence alignment'', discovering locations with similar substrings.

\subsection{Genomics and Acceleration}

Since genome analysis is becoming essential in various fields of human life, while the price of getting genomic sequence data has been decreasing, it is becoming extremely difficult to analyze the enormous size of the genome sequence data without acceleration.
Much prior research has introduced acceleration to sequence alignment, read mapping, classification, and structure similarities for faster analysis via various acceleration platforms, such as GPUs~\cite{zeni2020logan,cheng2018bitmapper2,kobus2017accelerating,nishimura2017accelerating,wilton2015arioc,cadenelli2017accelerating}, FPGAs~\cite{liyanage2023efficient,li2021pipebsw,haghi2021fpga,qiao2019fpga,ham2020genesis}, ASICs~\cite{turakhia2019darwin,fujiki2018genax,huangfu2018radar,huangfu2022beacon,cali2022segram,cali2020genasm}, PIMs~\cite{zhang2023aligner,jahshan2023dash,zou2022biohd,shahroodi2022krakenonmem,chen2023gem,kaplan2020bioseal}, Quantum computing~\cite{varsamis2023quantum}, and In-Storage processing~\cite{mansouri2022genstore}.

Among the acceleration platforms, GPUs and FPGAs are the most popular platforms because they are more affordable and off-the-shelf available.
Thanks to the capability to compose fine-grained application-specific reconfigurable logic, FPGAs often show better performance and power efficiency than GPU-based accelerators, especially for the representative genome analysis algorithm, sequence alignment~\cite{cheng24fpga,guo2019fpga}.

\subsection{Memory and Storage Systems}

Along with the looming end of Moore's law scalability, the future of memory scalability is also uncertain.
Density of capacitor-based DRAM in the 2D space is expected to stop scaling, with no feasible alternative~\cite{endofdram}.
One direction of continued improvement is through 3D stacking, for memories such as High Bandwidth Memory (HBM), where bandwidth continues to scale with ever-wider communication bandwidth using Through-Silicon Vias (TSV).
However, while bandwidth continues to scale, the height of 3D stacking faces physical limitations, making capacity scaling more uncertain~\cite{kim2015architectural,shiratake2020scaling,kim2024present,hyun2024pathfinding}.
This is a concerning prospect as the capacity scale of problems such as computational genomics continues to grow.
As a result, heterogeneous systems combining DRAM, NVMe, and more are under active investigation~\cite{ziegler2022scalestore,cadenelli2017accelerating}, as well as memory disaggregated over a network, using emerging fabrics such as CXL~\cite{gouk2023memorycxl,huangfu2022beacon}.

\subsection{Genome Compression}

As the scale of genomic data continues to grow, compression is becoming more important to reduce both data archiving and transmission.
Unfortunately, conventional data-agnostic compression algorithms dictionaries and Huffman coding have shown very little compression efficiency due to the high entropy of genomic data~\cite{kokot2022colord,mansouri2020newdnacompression}.

Many genome-specific compression schemes are under active investigation, and they can largely be categorized into two groups.
The first group is reference-free compression, where the algorithm only looks at the genome to compress, and takes advantage of statistical trends it can find~\cite{chandak2018spring,kokot2022colord}.
These schemes typically require little memory, but achieve low compression efficiency.
The second group is reference-based compression, which uses a reference, (or ``consensus'') and takes advantage of the similarity between individuals of the same species to encode only the differences~\cite{bonfield2022cram,kokot2022colord}.
These schemes often achieve very high compression ratios in the hundreds, but require more memory because the entire reference needs to be memory-resident.
Other approaches are being explored as well, including using machine learning models ~\cite{silva2020efficientdnann,cui2020compressinggenomedeep,sheena2024gencoder}.

Acceleration of genome compression is also under active investigation to overcome the slow speed of these algorithms, using FPGAs~\cite{arram2015fpgareferencecompressiongenomic,chen2023efficientsequencingcompressionfpga}, and GPUs~\cite{guo2013gpu,amich2020gpu,deluca2022gpu}.

%% file: compression.tex
\section{\name{} Compression Algorithm}
\label{sec:algorithm}

The key novelty of \name{} is its compression algorithm, a reference-based lossless algorithm co-optimized with hardware acceleration to achieve high performance for both compression and decompression, without sacrificing compression efficiency.
The primary goal of the design is to achieve a high enough compression ratio and decompression performance in hardware to close the performance gap between PCIe and on-device DRAM, providing \name{} with practically unlimited amounts of high-performance memory.
Both compression and decompression performance are important goals, for analytics acceleration as well as archiving and distribution workloads.

\subsection{Algorithm and Data Structures}

The \name{} compression algorithm belongs to the class of reference-based compression algorithms, which find matching regions between the input ``\emph{target}'' data and a ``\emph{reference}'' genome.
\name{} addresses two key limitations to compression and decompression performance: Speed of discovering matching regions during compression, and complexity of decoding variable-width encoded compression schemes.

To overcome these limitations, \name{} introduces the following three strategies.
Each strategy will be described in more detail, along with the compression and decompression algorithms below.

\begin{itemize}
    \item \textbf{Strategy~1}: Fixed-k k-mers and fixed-stride matching.
    \item \textbf{Strategy~2}: Fixed-width grouped header encoding.
    \item \textbf{Strategy~3}: Cuckoo hashes for match discovery.
\end{itemize}

Strategies 1 and 2 work together to improve decompression performance and resource efficiency, while strategies 1 and 3 work together to improve compression performance.
We demonstrate the minimal impact of these strategies on compression efficiency, in Section~\ref{sec:compression_efficiency}.

\subsection{Compression Algorithm}

\begin{figure}[htb]
    \centering
    \includegraphics[width=\columnwidth,page=4]{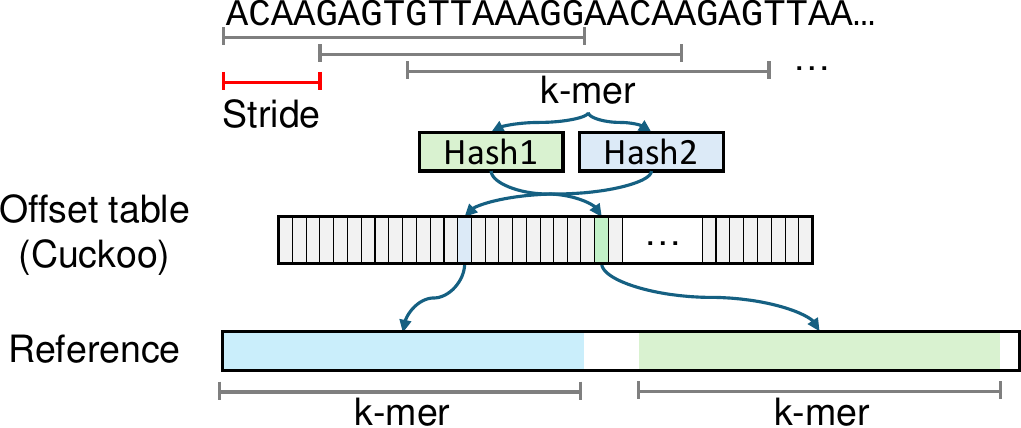}
    \caption{Reference matching with fixed-k, fixed-stride.}
    \label{fig:fixed-width}
\end{figure}

Figure~\ref{fig:fixed-width} illustrates how \name{} performs alignment between the target and reference with high performance.
The algorithm has two parameters: the width of the k-mer $K$, and the stride $S$.
Similar to how seeding works in modern sequence alignment algorithms, \name{} finds exact matches between the target and reference sequences in k-mer units.
Since we are looking for exact matches, we avoid using a complex index structure in favor of an $O(1)$ hash table, which is pre-constructed from the reference genome to hold the offset of each k-mer within the reference.
Such efficient lookup vastly improves \name{} performance compared to conventional schemes using suffix arrays or alignment algorithms~\cite{chen2023efficientsequencingcompressionfpga,arram2015fpgareferencecompressiongenomic}.
Since the offset table is constructed offline once for each reference and used repeatedly for the same organism, we adopt the cuckoo hash for better space efficiency.
Cuckoo hashes use two hash functions per query, so each query can have a choice of avoiding conflicts by selecting one of two locations.
Its algorithm ensures that all keys can be accommodated without conflict misses if the capacity of the table is 2$\times$ the set size of keys~\cite{pagh2004cuckoo}.

Each potential match from the reference is compared against the target k-mer.
If a match exists, the whole k-mer is encoded compactly as a single offset, and the window moves forward by $K$ instead of $S$.
If neither is a match, then the mismatched stride is encoded verbatim, and the window moves forward by $S$ to check the next k-mer. 
In our eventual prototype, $K$ was set to 64, and $S$ was set to 16. 
Only considering exact matches and having a wide stride can potentially trade off compression efficiency for higher performance.
The compression efficiency impact of these parameters is presented in Section~\ref{sec:compression_efficiency}.
We also note that \name{} actually uses four hash functions during compression, although the cuckoo hash is constructed using two.
This is because we also look for reverse complemented versions of the target k-mers, as sequencing machines emit both directions of reads.

\begin{figure}[htb]
    \centering
    \includegraphics[width=.7\columnwidth,page=1]{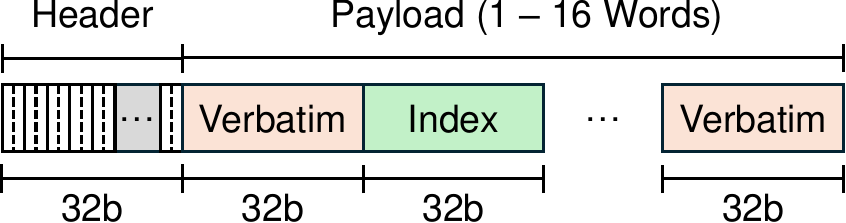}
    \caption{Grouped header compression encoding.}
    \label{fig:compressed_format}
\end{figure}

Our choice of $S$ and $K$ enables high performance and resource efficiency beyond just a high compression ratio.
Figure~\ref{fig:compressed_format} shows the compressed encoding scheme.
Firstly, \name{} uses a grouped header scheme to enable fast decompression, similar to Group VarInt~\cite{dean2009challengesgroupvarint}.
Second, since the 2-bit encoded human reference is less than 4~GB (\textasciitilde 700~MB) and $S=16$ means verbatim encoding of a stride takes up 32 bits, every field in the encoding scheme has a fixed width of 32 bits.
This scheme enables extraordinarily simple decoding schemes, as described in Section~\ref{sec:decompression_arch}.

The header for each stride consists of two bits, meaning the payload for the grouped encoding scheme can consist of up to 16 32-bit words, either verbatim encodings or indices.
The payload can have fewer words, since \name{} does not encode the index for ``\emph{continuations}'', which are matches that occur immediately after the previous match in terms of the offset within the reference.
Instead, continuations are only specified in the header.
Specifically, \texttt{00} specifies verbatim encoding, \texttt{01} specifies a forward match, \texttt{10} specifies a reverse complement match, and \texttt{11} specifies a continuation without a payload.

The hardware-accelerated implementation described in Section~\ref{sec:compression_arch} also introduces transparent performance enhancements, including probabilistic filtering, to overcome the computation and memory system bottleneck of this algorithm during compression.

\subsection{Decompression Algorithm}

Thanks to the hardware-aware considerations during encoding, the decompression process is very simple.
The decompression software or accelerator simply scans through the header, and decodes the 32-bit words in the payload as verbatim encodings, or as indices to look up the reference genome.
If the next element is a continuation, depending on whether the last match or continuation was in forward or reverse order, the next or previous k-mer is fetched from the reference genome.

We show in Section~\ref{sec:decompression_arch} that the grouped header encoding allows very efficient decoding of both verbatim values and efficient reference genome lookups.

%% file: software.tex
\section{Software Manager and Interface}
\label{sec:software}

The software interface of \name{} wraps around both the Xilinx XRT environment and the compression functionalities, to provide a familiar and transparent interface to acceleration on both compressed and uncompressed data.

\subsection{Programming Interface}

The \name{} software interface provides a familiar, XRT-style wrapper around accelerated kernel calls.
Let's consider a kernel \texttt{accel} with two device-side buffers ``\texttt{a}'' and ``\texttt{b}'' as parameters, where ``\texttt{a}'' is a genome data source.
If ``\texttt{a}'' comes from a \name{}-managed compressed source, subsections of the file can be transmitted as part of the kernel execution simply by setting \texttt{a.source = \name{}.compressed\_file(1); a.offset = N; a.size = M;}, and then passing it to the kernel via \texttt{accel.kernel(a,b)}.

The current version of the \name{} decompressor only supports stream sources, meaning transmitted buffers are decompressed immediately and provided to the kernel over a FIFO interface.
If the kernel requires random access into a buffer, it must copy the decompressed data to a separate in-memory buffer.
We are working on building an on-device index for compressed data to overcome this limitation.

\subsection{Index Structure}

One of the most important features of compressed data management is random access.
Downstream processing, such as graph construction based on reads, requires each read to be accessed separately, and the fundamental workload of reference-based alignment requires random access capabilities into the reference.

\name{} achieves this feature using a B+tree data structure, as hinted at in Figure~\ref{fig:overall}.
During compression, a B+tree is constructed with the file-internal offset as the key.
The unit element of insertion and lookup is a chunk of compressed data, which shares a single 32-bit header.
While the size of a decompressed chunk can be up to 128 bytes large in our prototype, we discovered that the random access requirement is typically coarser than this, in the unit of long reads.

\subsection{Reference Genome Encoding}

The software environment also must store the reference genome used for compression, because the compression process is split between hardware and software, as described in Section~\ref{sec:compression_arch}.
As described in Section~\ref{sec:compression_arch}, the hardware portion of compression is primarily responsible for calculating the hash function.
On the other hand, the software must perform a lookup into the cuckoo hash table, which is too large to store comfortably in the accelerator, and discover the correct cuckoo hash slot (if any) by comparing the target k-mer against up to four k-mer substrings sampled from the reference genome according to the cuckoo hash lookup.

Because \name{} stores the reference in a compact 2-bit encoded format, comparing k-mers can cause performance overhead due to sub-byte addressing.
For example, if a k-mer starts from offset 7, with two bit encoding, the 7th base starts in bit 6 of byte 2.
Comparing k-mers in this setting requires repeated shifting operations, which can have performance overhead.
To overcome this, \name{} stores four copies of the reference, each shifted from the original by 2 to 6 bits.
During compression, one of the copies are chosen according to the ``offset mod 4'' value, and then fast \texttt{memcmp} can be used since the string would be aligned along byte boundaries.

Both alternatives to storing four binary copies: Storing byte-alined ASCII files, and performing shifts on the fly, showed performance degradation by 4$\times$ on average.

%% file: hardware.tex
\section{Acceleration Platform and Interface}
\label{sec:hardware}

\begin{figure}[htb]
    \centering
    \includegraphics[width=.65\columnwidth,page=5]{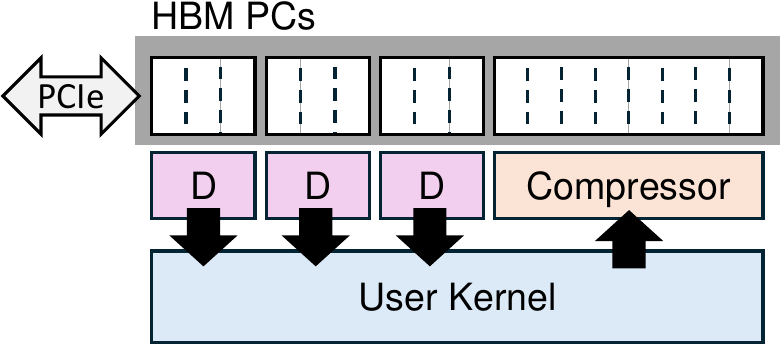}
    \caption{An example design with three decompressors (D) and one compressor, consuming 17 HBM PCs.}
    \label{fig:hardware}
\end{figure}

As illustrated in Figure~\ref{fig:overall}, \name{} provides the user hardware accelerator with a wrapper around the large genomic dataset.
The primary hardware components in the architecture are the decompression accelerators and the compression accelerators, illustrated with an example configuration in Figure~\ref{fig:hardware}.
Each compression and decompression accelerator needs access to a fixed-size buffer in HBM, so they must be coupled with sets of HBM pseudo channels (PC).
This is described in more detail in the following sections.

For an FPGA deployment of \name{}, the number of compressors and decompressors in the system can be configured for each application's requirements, within the physical boundaries of the platform.
For example, the prototype's U50 platform has 32 HBM channels, 256~MB each, and the human reference genome requires the capacity of three channels.
As a result, up to 10 decompressors can be instantiated.

\subsection{Compression Accelerator}
\label{sec:compression_arch}

Due to the memory capacity requirement of the cuckoo hash offset table, \name{} divides the compression process between hardware acceleration and the software manager.
For example, the cuckoo hash table to encode the human genome is at least 16~GBs, leaving no space even on medium-range FPGA platforms such as Alveo U55C.
Instead, it is left to the software to maintain and query the offset table in memory, perform the reference comparisons, and encode the compressed data, as described in Figure~\ref{fig:fixed-width}.
The hardware accelerator is responsible for providing a stream of hash values and supplementary information to the software over PCIe.
The hardware component is mainly responsible for two tasks: Computing the hash values for the cuckoo hash, and performing probabilistic filtering to reduce the load on the software.

\begin{figure}[htb]
    \centering
    \includegraphics[width=\columnwidth,page=6]{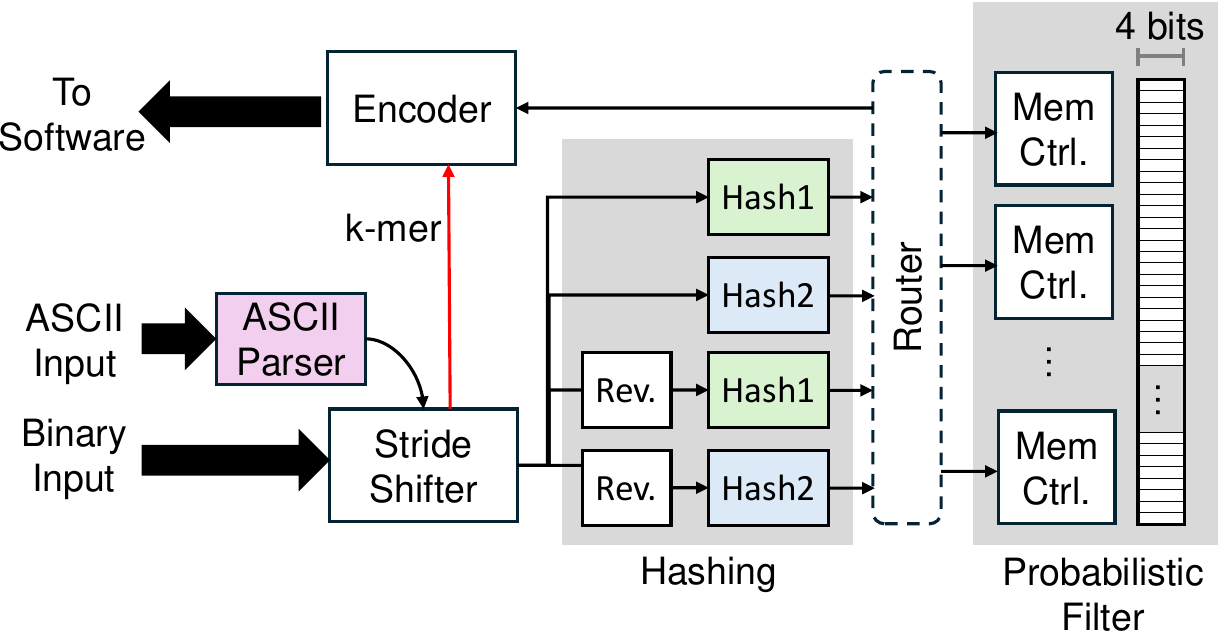}
    \caption{Hardware portion of the compressor.}
    \label{fig:compression_hardware}
\end{figure}

Figure~\ref{fig:compression_hardware} shows the architecture of the hardware accelerator.
The accelerator ingests input data either in the form of ASCII-encoded text such as FASTA, or as a 2-bit binary encoded format generated by user kernels.
ASCII input is first encoded in binary format, and a stride shifter emits a stream of k-mers into an array of four pipelined hash function modules.
Since our prototype uses a $K$ value of 64, the datapath between the stride shifter and the four hash functions is 128 bits wide.
The hardware needs four hash functions because two hashes are needed to query the cuckoo hash, and two more are needed to query the cuckoo hash based on the reverse complement of the k-mer.
The computed hash values are independently routed to one of many memory controllers, each interfacing with an HBM PC, to check the probabilistic filter.
The filter results, along with the original k-mer as emitted by the stride shifter, are collected at the outbound encoder to organize them into a software-friendly format.
By default, \name{} uses the Murmur3 hash function to achieve both high software and hardware throughput.

The probabilistic filter is not part of the algorithm as described in Section~\ref{sec:algorithm}, but a transparent optimization to improve effective software performance.
Its task is to reduce the number of hash values the software has to verify.
This is done via another, much smaller cuckoo hash table in accelerator memory, as depicted in Figure~\ref{fig:compression_hardware}.
The four hash values point to the same index as they would in the software-side hash table, but each element in the accelerator table is only 4 bits long.
Instead of storing the 4-byte offset into the reference, the 4 bits hold a substring of the reference itself.
Specifically, it holds the least significant 4 bits of the reference k-mer which maps to this location.
More specifically, given a k-mer ``\texttt{str}'', which generates four hash values \texttt{hash$_1$} to \texttt{hash$_4$}, if \texttt{str[3:0]} does not match \texttt{filter\_table[hash$_1$]}, then we can be sure that software does not need to check \texttt{hash$_1$}.
The performance impact of this optimization is quite significant, as presented in Section~\ref{sec:compression_performance}.
As a result, the compression accelerator needs about 2~GB of HBM memory, as depicted in Figure~\ref{fig:hardware}.

Because 2~GB of HBM memory is spread across multiple pseudo channels (256~MB each in our prototype), which can operate in parallel to achieve very high aggregate bandwidth, one may worry that the encoder and its datapath may not be fast enough to keep up with memory bandwidth.
Unfortunately, the inherently random, fine-grained 4-bit table lookups result in very low effective memory performance.
Not only is random access inherently slower on DRAM technologies, including HBM~\cite{hbm_xilinx}, but the minimum burst length of HBM is typically long, around 256 bits, resulting in severe I/O amplification.
This is one of the reasons hash table lookups need to be filtered using fast accelerator HBM, instead of the host software working with larger, but slower DDR memories.

The software needs to compare the target k-mer with the potential reference k-mers discovered via the cuckoo hash, but the software may not have access to the target k-mers.
This may be because genomic data is being generated by the user kernel, or because the target exists in ASCII format, which is computationally expensive to compare against, which we demonstrate in Section~\ref{sec:compression_performance}.
To overcome this limitation, the hardware sends both the target k-mer, along with the four hashes to the software for comparison.
Since the k-mer for the last shift would exist in software already, each shift only needs to send the next 32-bit stride.
In our prototype configuration of $K=64$, this means each stride shift results in 20 bytes of data transfer (4 bytes $\times$4 hash values + 4-byte stride).

The remaining process of compression and data organization in software is described in more detail in Section~\ref{sec:software}.

\subsection{Decompression Accelerator}
\label{sec:decompression_arch}

\begin{figure}[htb]
    \centering
    \includegraphics[width=\columnwidth,page=7]{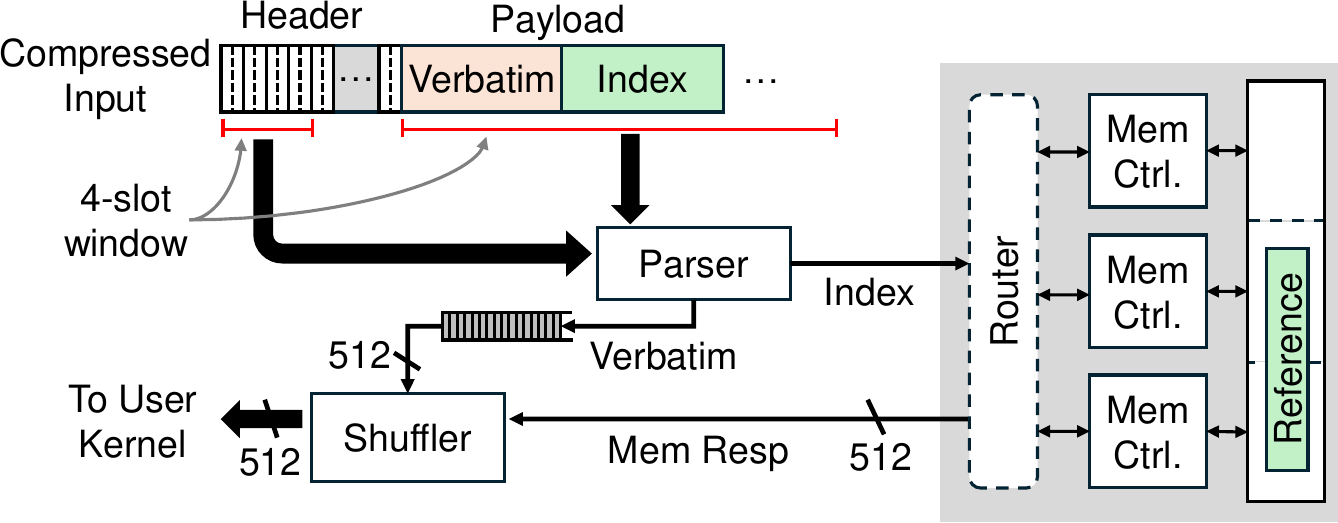}
    \caption{Decompression hardware accelerator.}
    \label{fig:decompression_hardware}
\end{figure}

The architecture of each decompression accelerator is very simple despite the high performance, thanks to the algorithmic modifications that simplify hardware design.
Figure~\ref{fig:decompression_hardware} presents the architecture of a compression accelerator.
The accelerator scans through both the header and payload of the compressed input using a sliding window of width 4.
The accelerator needs to consider a window of slots at once, because the k-mer loaded from memory for each encoded index is only 64 bases, or 128 bits.
Sending a memory request for each index would severely degrade performance if the memory bus is wider than this.
For example, our FPGA prototype running at 250~MHz needs a 512-bit memory bus to make the best use of memory bandwidth, and transmitting 128-bits per cycle will waste too much bandwidth.
Similarly, verbatim encoded data is stored in units of 32 bits, and decoding one such element per cycle will lead to bandwidth waste.

The sliding window parser considers four header bits at once, and if it discovers consecutive verbatim encodings, or consecutive continuations, it can either decode up to four verbatim encodings in one cycle, or issue a single 512-bit memory request instead of four requests over four cycles.
We emphasize that this is made possible thanks to both strategies 1 and 2 in Section~\ref{sec:algorithm}, fixed-width encodings and grouped headers.
Having grouped headers allows the parser to quickly determine consecutive elements, and fixed encoding width not only makes headers small since size information does not need to be encoded, but it also makes it simple to slide a fixed-width window through both header and payload.

On the other hand, if each header element was interleaved with variable-width payload elements, the next header can only be decoded \emph{after} the current payload was decoded, making it difficult to scan a wide window at once.
Furthermore, looking at the next element also would require variable-width shifting, which is known to be resource-heavy and slow.

Memory read responses and the parsed verbatim elements are collected at the shuffler, which uses a multi-cycle pipelined shifter and a compaction module based on sorting networks to shift and concatenate data across different cycles and remove gaps.
The resulting output is a gap-free continuous stream of genomic data, which is naturally expected by the user kernel.

We note that although three HBM channels (250~MB each) are assigned to each decompressor to accommodate the whole human reference genome, the single 512-bit output bus of the decompressor never seems to become a bottleneck.
On the one hand, three memory channels operating at peak sequential bandwidth will overrun the decompressor's output by almost a factor of three.
However, we experienced with our FPGA prototype implementation that the random nature of the reference lookups results in low effective bandwidth of each channel, resulting in a good balance between the bus widths.
For this reason, the reference stored in HBM for one decompressor cannot be shared with another decompressor.
Since memory bandwidth is already being fully utilized, two decompressors sharing a reference memory will cause immediate performance degradation.
More detailed insight into bandwidth usage is provided in Section~\ref{sec:compression_performance}.

%% file: evaluation.tex
\section{Evaluation}
\label{sec:evaluation}

We evaluated \name{} using a prototype constructed using an affordable Xilinx U50 Datacenter FPGA accelerator card with 8~GB of 3D-stacked HBM, plugged into either a desktop-class or server-class computers with up to 48 Xeon cores and hundreds of GB of DDR4 memory.
The U50 FPGA card is plugged into the host machine via a PCIe Gen3x16 connection, where our framework was able to achieve a steady-state transfer rate of 9.1 GB/s download and 8.4~GB/s upload.

\subsection{Resource Utilization}

Table~\ref{tab:resource_utilization} presents the resource utilization of the \name{} compressor and decompressor modules.
These resources are needed in addition to the baseline utilization by the Xilinx XRT platform.
Because both modules depend on in-memory data structures for reference and probabilistic filtering, the primary resource limitation is HBM.
We demonstrate below that \name{} can achieve memory-class performance even within the stringent resource limitations of the affordable U50 device.

\begin{table}[htb]
    \centering
    \caption{Resource Utilization on the U50 FPGA.}
    \label{tab:resource_utilization}
    \begin{tabular}{|l||c|c|c|}
        \hline
        Module & LUT & DSP & HBM Channels \\
        \hline
         Compressor $\times$1 &  36K (4.14\%) & 480 (8\%) & 8 (25\%) \\
         Decompressor $\times$1 & 26K (2.99\%) & 4 (0.07\%) & 3 (9.3\%) \\
        \hline
    \end{tabular}
\end{table}

\subsection{Evaluated Datasets}

Table~\ref{tab:evaluated_datasets} presents the datasets we have used to evaluate \name{}, spanning reference and long-read datasets from human, mouse, and corn genomes.
We note that we also made use of three additional reference genomes, HG19~\cite{raney2024ucsc}, GRCm39~\cite{GRCm39}, and B73~\cite{B73}.
These were used as compression references, and not included in the analysis. 

\begin{table}[htb]
    \centering
    \caption{Evaluated datasets}
    \label{tab:evaluated_datasets}
    \begin{tabular}{|l|c|c|c|c}
        \hline
        Dataset & Organism & Type & Size (GB) \\
        \hline
        \hline
         HG16~\cite{raney2024ucsc} & Human & Reference & 2.67 \\
         HG38~\cite{raney2024ucsc} & Human & Reference & 2.75 \\
         \hline
         HG002~\cite{HG002} & Human & Long-Read & 121.19 \\
         HG003~\cite{HG003} & Human & Long-Read & 204.78 \\
         HG004~\cite{HG004} & Human & Long-Read & 205.62 \\
         mouse-rep1~\cite{mouse-rep1} & Mouse & Long-Read & 129.85 \\
         mouse-rep2~\cite{mouse-rep2} & Mouse & Long-Read & 131.48 \\
         maize-B73-rep1~\cite{maize-B73-rep1} & Maize & Long-Read & 130.77 \\
        \hline
    \end{tabular}
\end{table}

\subsection{Compression Efficiency Evaluation}
\label{sec:compression_efficiency}

We first present the compression efficiency of \name{}'s hardware-optimized genome compression algorithm.
The algorithm is defined with two parameters: stride $S$ and K-mer width $K$.
We performed exhaustive problem space exploration to determine the optimal $K$ and $S$ parameters, which we summarize below.
We note all compression ratios are based on the ASCII text (8 bits per base)  representation.

\subsubsection{Impact of stride width}
\begin{figure}[htb]
    \centering
    \includegraphics[width=\columnwidth,page=1]{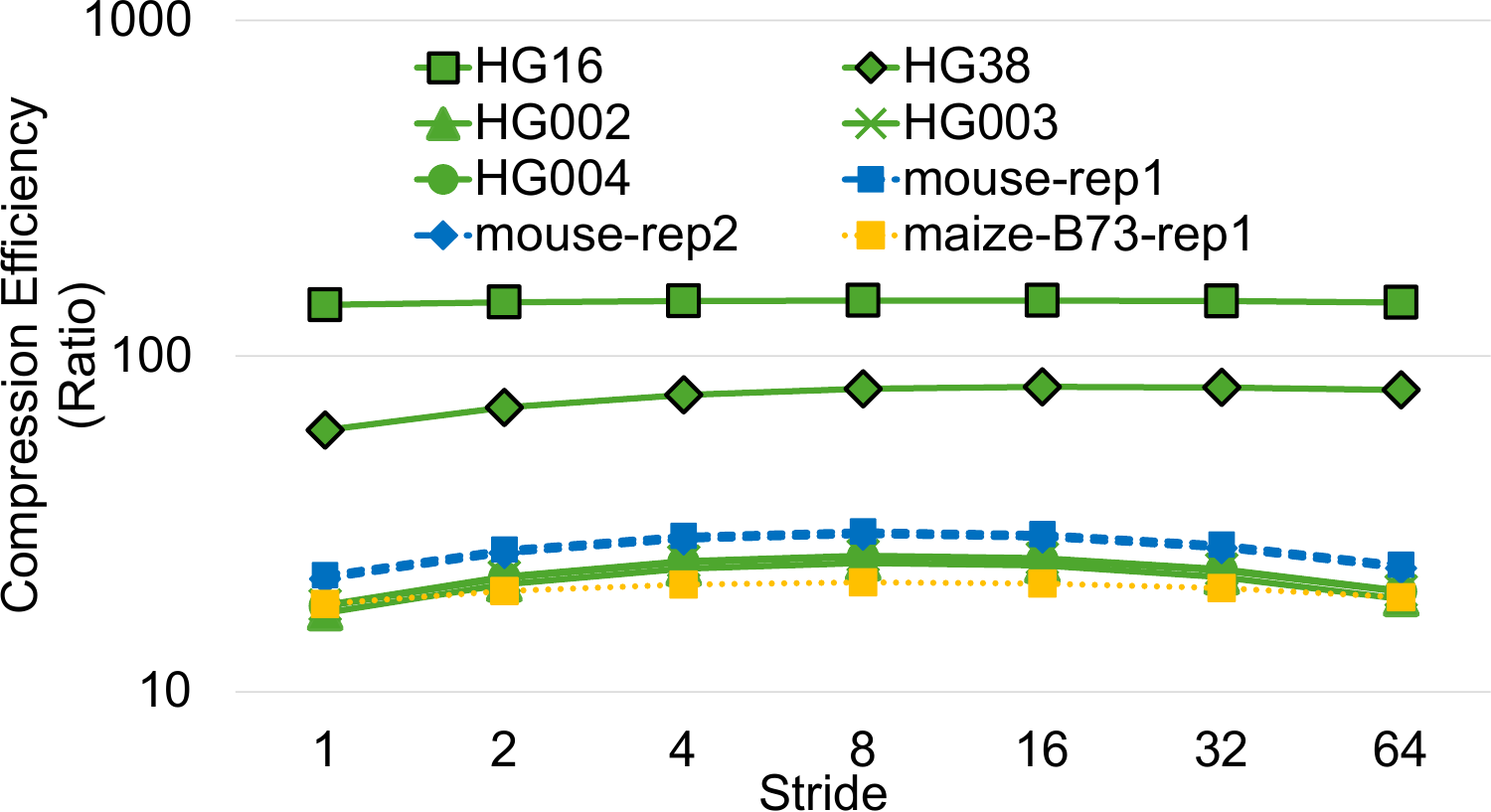}
    \caption{Compression efficiency with different $S$ (Log-scale).}
    \label{fig:compression_efficiency_stride}
\end{figure}

Figure~\ref{fig:compression_efficiency_stride} represents the compression efficiency with different stride configurations, for different datasets.
Each dataset uses a reference genome appropriate for its organism.
$K$ is fixed to 64 in this presentation, but similar behavior was observed with different $K$ values.
Measurements show that $S$ values of 8 and 16 yield the best compression efficiencies.
\name{} uses 16 by default for $S$, since the stride of 16 results in 32-bit verbatim encoding, which is the same width as the reference genome index for matches.
This makes the hardware decoder design more compact.

The reason smaller $S$ values result in lower compression is due to the header overhead.
While smaller strides result in slightly more k-mers being matched to the reference, it also results in more header bits because each verbatim encoding is smaller.

\subsubsection{Impact of $K$ width}
\begin{figure}[htb]
    \centering
    \includegraphics[width=\columnwidth,page=2]{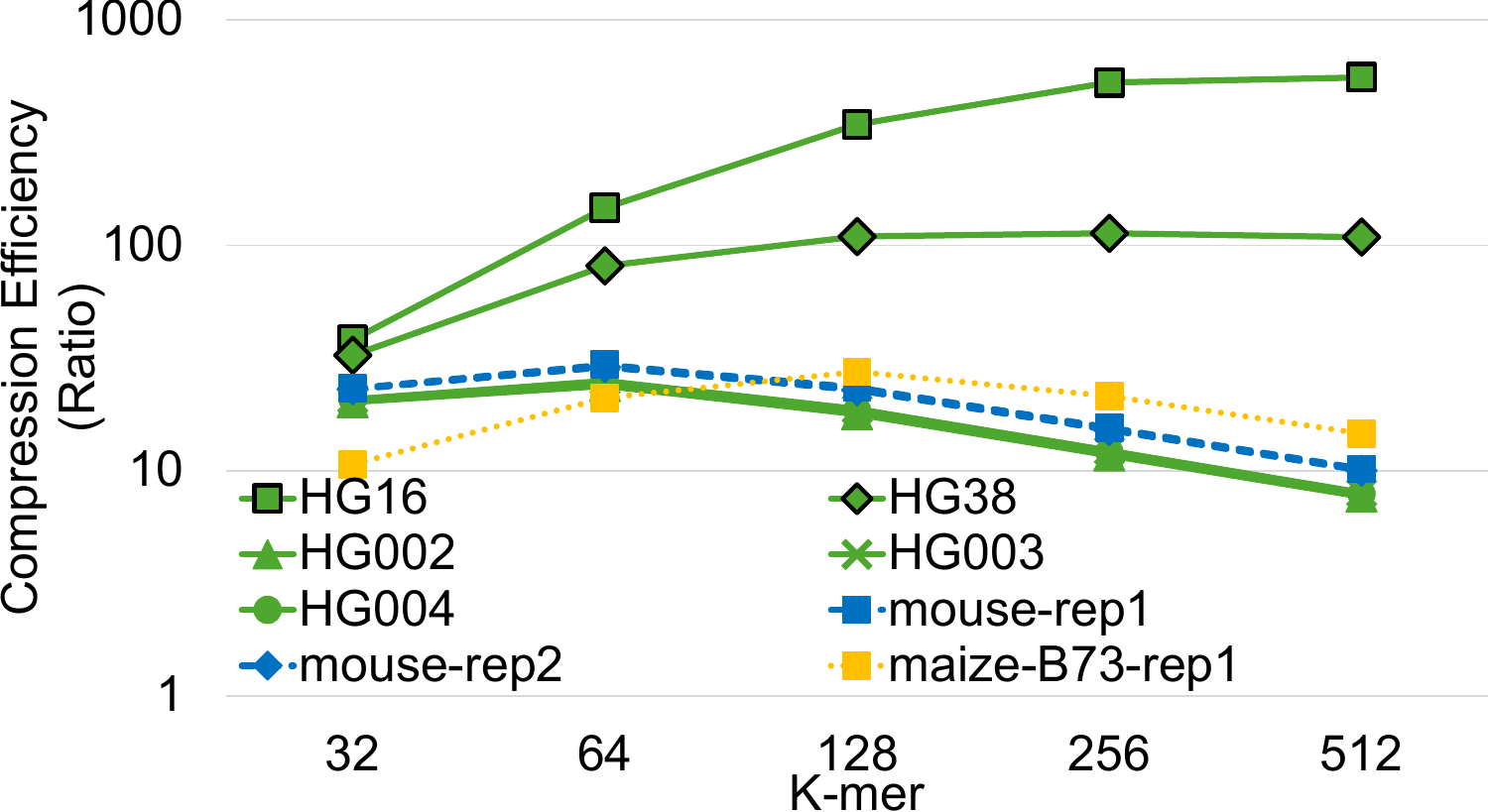}
    \caption{Compression efficiency with different $K$ (Log-scale).}
    \label{fig:compression_efficiency_kmer}
\end{figure}

Figure~\ref{fig:compression_efficiency_kmer} presents the compression efficiency of \name{} with different $K$ values.
$S$ is fixed to 16 in these experiments.
The results fall into two different categories.
When compressing reference genomes, (HG16 and HG38) larger $K$ typically yields better compression.
For long-read sequences, compression efficiency peaks at 64 and then declines.
This difference appears to come from quality differences between the two categories.
The pre-assembled references have most read errors and other noise removed, and comparison between them can take advantage of the genetic similarity between the two individuals.
On the other hand, long-read datasets still have numerous read errors and other noises, which makes long matches less likely.

Most experiments with long reads show the best compression at $K$ of 64 (Human, Mouse) and 128 (Maize).
We choose the $K$ value of 64 by default, to emphasize its use for Humans.

\subsubsection{Comparison against state-of-the-art}
\begin{figure}[htb]
    \centering
    \includegraphics[width=\columnwidth,page=3]{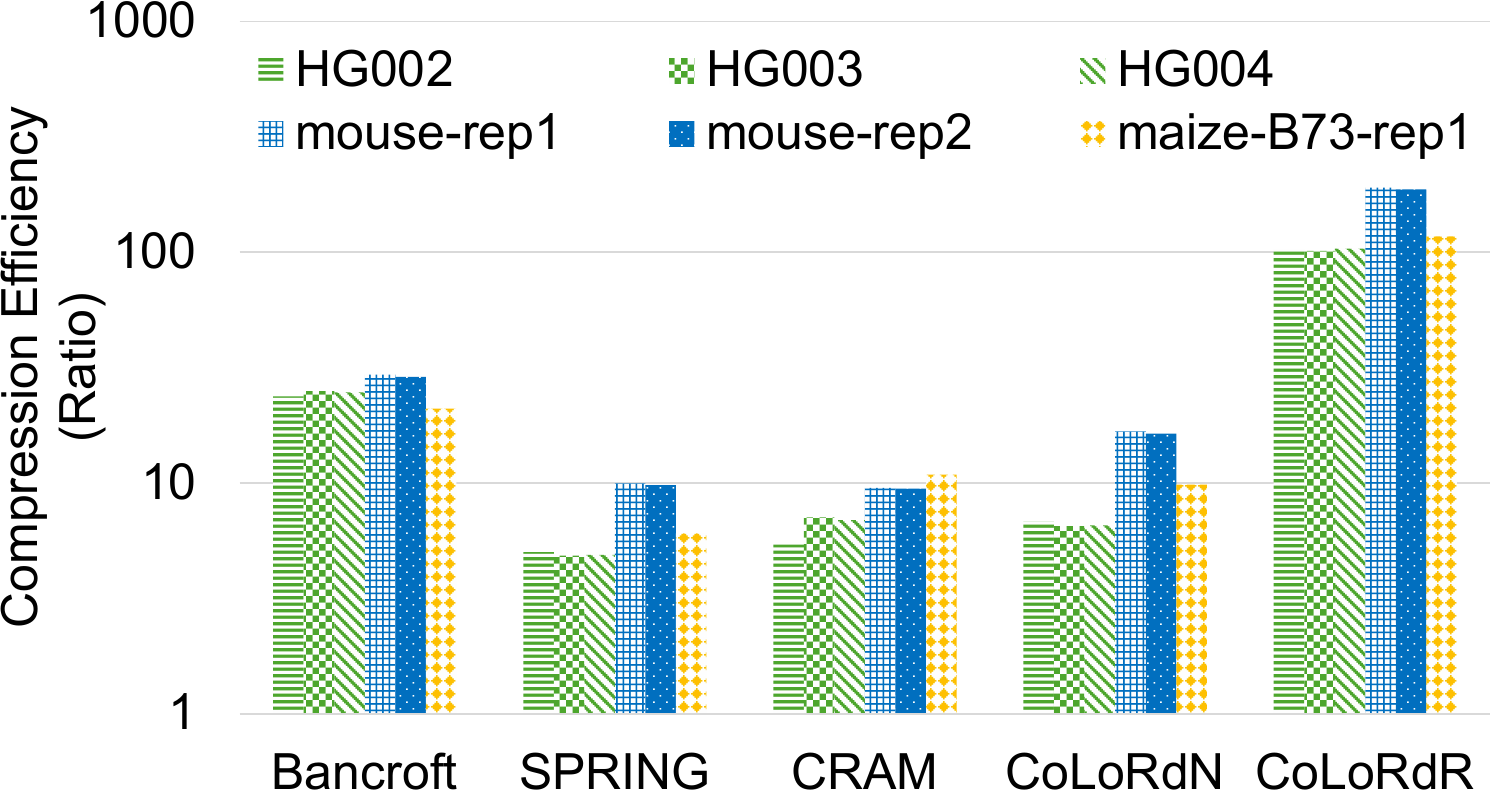}
    \caption{Compression efficiency comparison with state-of-the-art tools (Log-scale).}
    \label{fig:compression_efficiency_comparison}
\end{figure}

Figure \ref{fig:compression_efficiency_comparison} evaluates the compression efficiency of \name{} against other state-of-the-art genome-specific compression tools.
SPRING (2018, \cite{chandak2018spring}) and CoLoRdN (2022, \cite{kokot2022colord}) are reference-free algorithms, while CRAM (2022, \cite{bonfield2022cram}) and CoLoRdR (2022, \cite{kokot2022colord}) are reference based.
\name{} was configured with the default parameter of $S=16$, $K=64$.
We emphasize that the y-axis is a log scale.
The \name{} compression algorithm, even with its optimizations geared towards efficient hardware implementations, achieves competitive efficiency compared to state-of-the-art algorithms.

\subsection{Compression/Decompression Performance}
\label{sec:compression_performance}

Here, we present the performance of the compression and decompression accelerators implementing the \name{} algorithm.

\subsubsection{Decompression performance}
\begin{figure}[htb]
    \centering
    \includegraphics[width=\columnwidth,page=4]{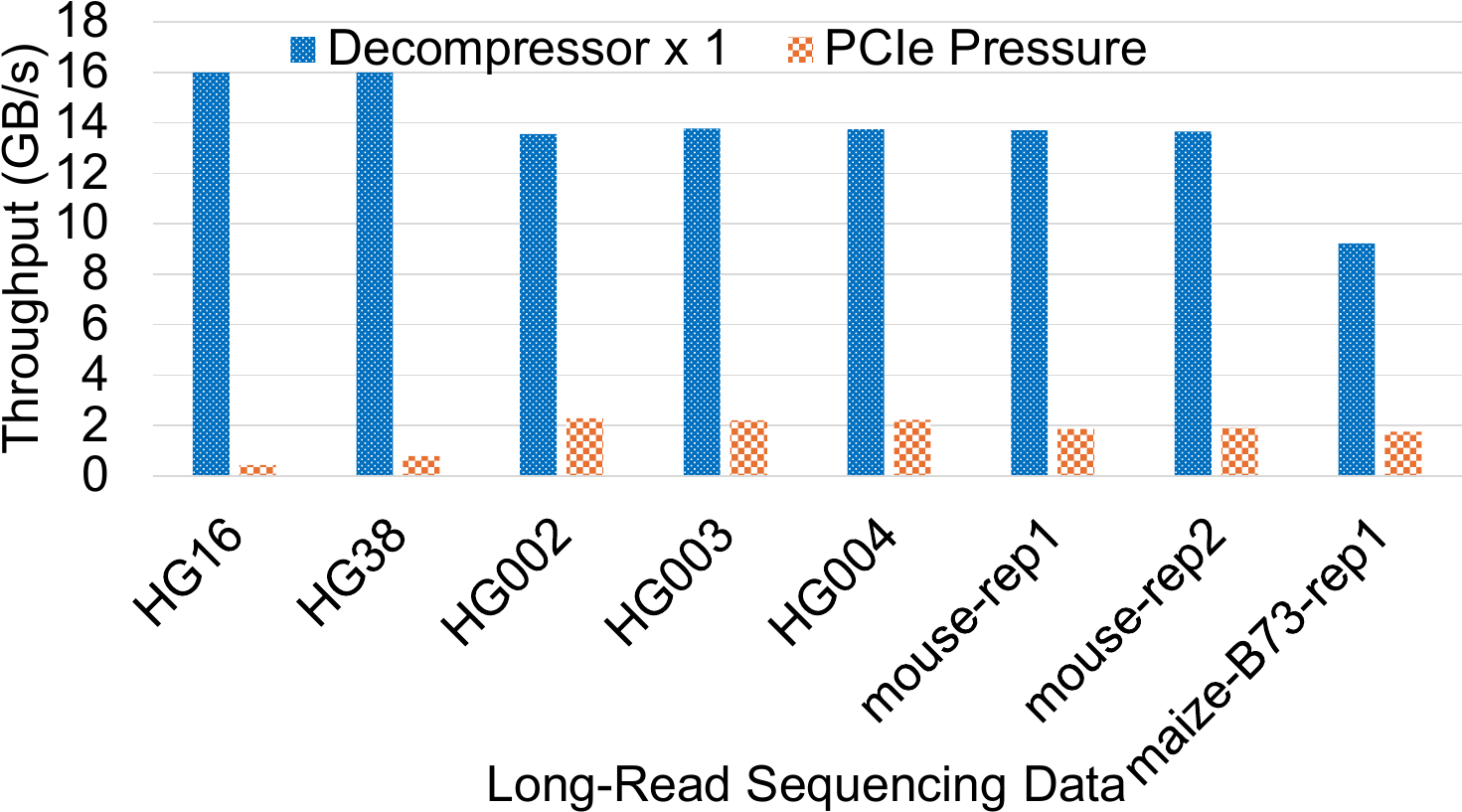}
    \caption{Single decompressor throughput.}
    \label{fig:single_decompressor_throughput}
\end{figure}

Figure~\ref{fig:single_decompressor_throughput} presents the throughput of a single decompressor module, with different datasets.
It also presents the bandwidth pressure put on the PCIe to achieve each decompressor output.
Thanks to the simple hardware design facilitated by the algorithmic optimizations, each decompressor can very nearly saturate the 512-bit output bus for all datasets tested, while consuming only 3\% of the chip space on the U50 FPGA.
Such high performance is also achieved while putting minimal bandwidth on the PCIe bus, thanks to the high compression efficiency.

\begin{figure}[htb]
    \centering
    \includegraphics[width=\columnwidth,page=5]{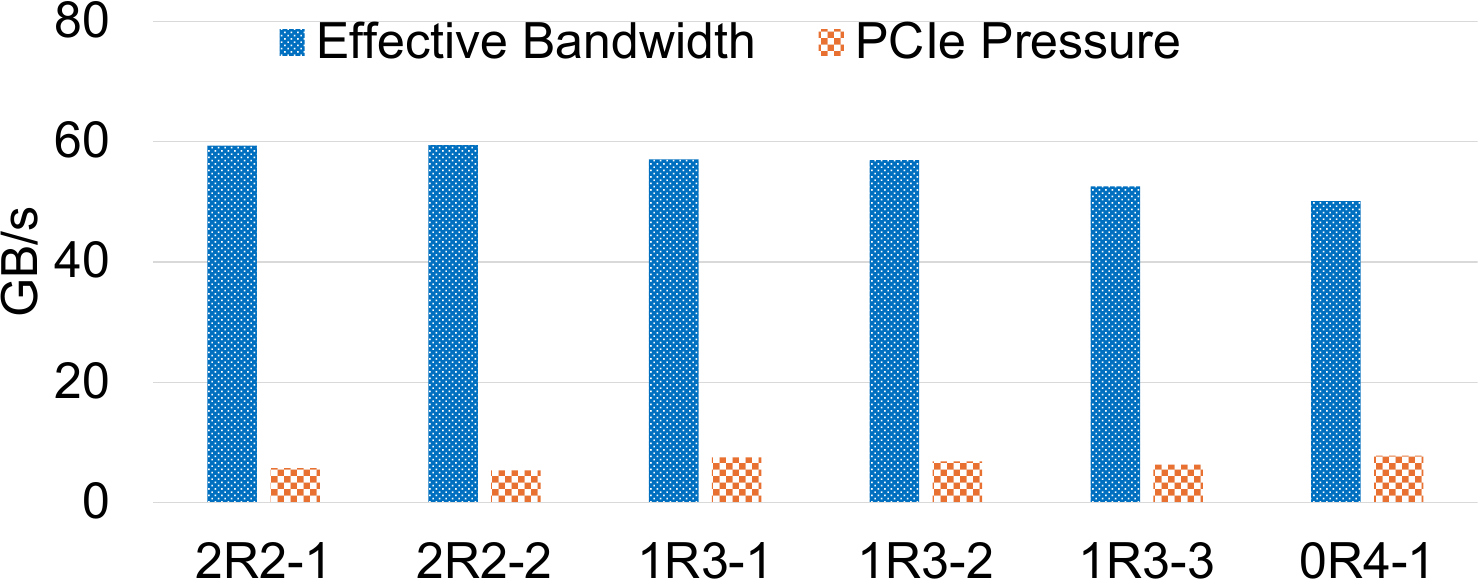}
    \caption{Ensemble decompressor throughput.}
    \label{fig:ensemble_decompressor_throughput}
\end{figure}

We also present the decompressor performance when working with realistic ensembles of genomic datasets.
Figure~\ref{fig:ensemble_decompressor_throughput} presents the total bandwidth achieved with \name{}, when working with multiple datasets in parallel through multiple decompressor channels.
The evaluated ensembles are described in Table~\ref{tab:decompressor_set_1}.
The names of each set describe how many references are used, and how many long read datasets are used (e.g., ``1R3'' means one reference and three long reads).
Each set represents a type of analysis.
For example, 2R2-1 represents a multi-reference alignment workload with a large read dataset, and 1R3-4 represents aligning a very large read dataset against a single reference, and 0R4-1 represents pairwise alignments for De Novo assembly.
We note that datasets for different organisms are not distinguished for this experiment.
Each dataset is used simply as a real-world dataset with different compression characteristics.

\begin{table}[htb]
    \centering
    \caption{The sets of decompressors}
    \label{tab:decompressor_set_1}
    \begin{tabular}{|l|c|}
        \hline
         Set & Composition \\
        \hline
         2R2-1 & HG16+HG38+HG002+HG003 \\
         2R2-2 & HG16+HG38+HG004+mouse-rep1 \\
         1R3-1 & HG38+HG002+HG003+HG004 \\
         1R3-2 & HG38+HG002+mouse-rep1+mouse-rep2 \\
         1R3-3 & HG38+mouse-rep1+mouse-rep2+maize-B73-rep1 \\
         0R4-1 & HG002+mouse-rep1+mouse-rep2+maize-B73-rep1 \\
        \hline
    \end{tabular}
\end{table}

Figure~\ref{fig:ensemble_decompressor_throughput} presents the total decompression performance across four decompressor modules, for each ensemble.
We can see that all configurations reach close to the peak bandwidth of four 512-bit buses (64~GB/s).
More importantly, none of the configurations reach the PCIe bandwidth bottleneck, as presented by the PCIe pressure bars.
0R4-1 comes close, but is still below the 8.4~GB/s upload limit.

We note that two copies of 1R3-1 or 0R4-1 are possible to deploy within the physical limitations of the U50, and such a configuration will overrun the PCIe bandwidth limitation, due to the relatively low compression efficiency.
But fortunately or unfortunately, the power throttling of the U50 HBM limits the performance before we run into this bottleneck, as presented in the following section.

\begin{figure}[htb]
    \centering
    \includegraphics[width=\columnwidth,page=13]{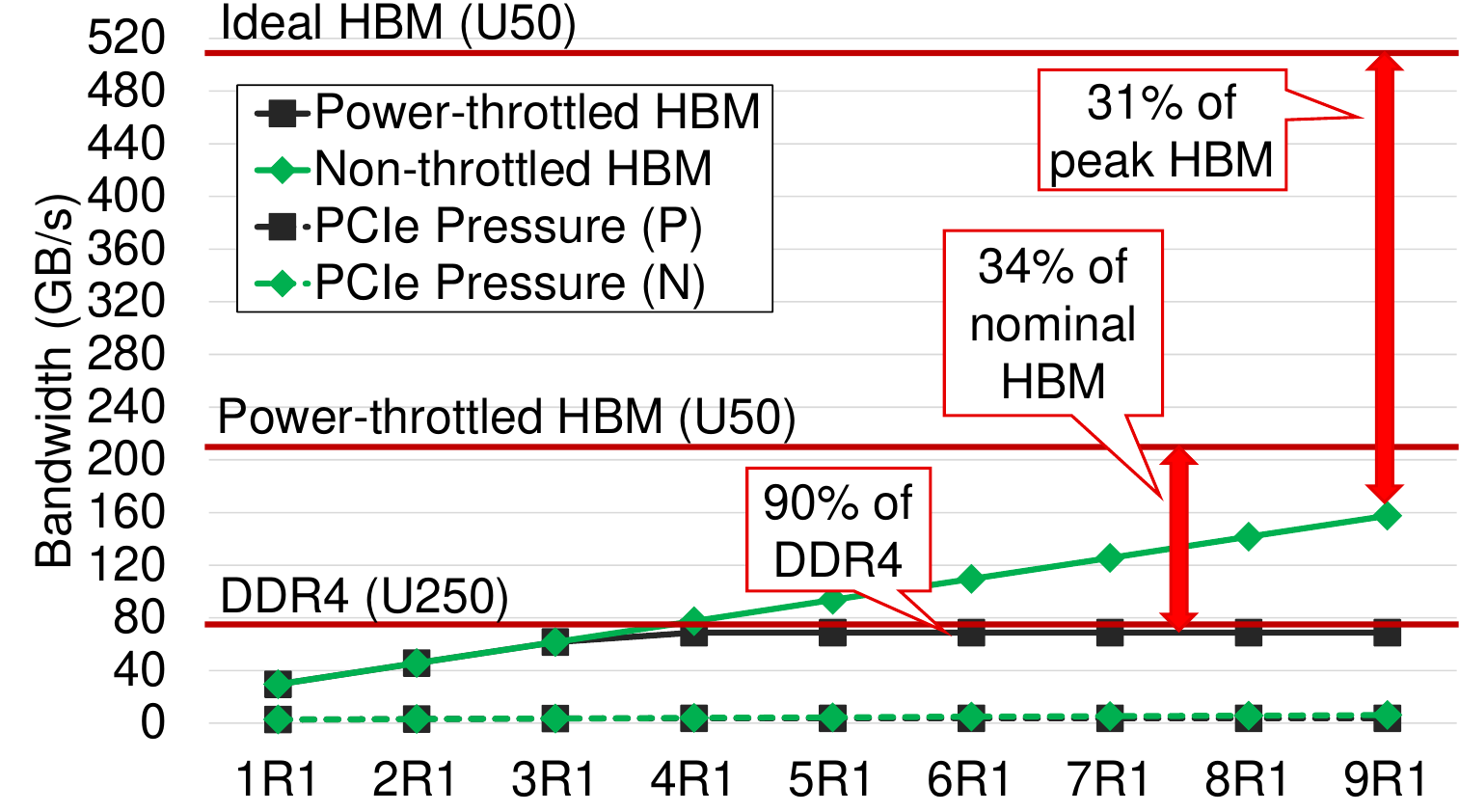}
    \caption{Impact of HBM power throttling on peak HBM throughput and \name{} performance.}
    \label{fig:throttled_bancroft}
\end{figure}

Figure~\ref{fig:throttled_bancroft} shows the best performance \name{} can achieve on the U50 platform, and compares it against other comparable configurations.
Here, we experiment with one long read dataset (HG002) aligned against multiple reference genomes, in a multi-reference alignment situation.
This was chosen as the best-case, but still realistic example, which can take advantage of the high compression efficiency on references to avoid the PCIe bandwidth limitation.
We notice that the real-world measured throughput of \name{} on the U50 stalls around 70~GB/s, due to the limited 7.8~W HBM power budget on the U50.
This is around 34\% of the nominal real-world performance of the U50 HBM, when the dataset is loaded in an uncompressed format directly from the HBM.
The measured performance of \name{} is similar to the peak memory bandwidth available on the higher-end Alveo U250 FPGA platform equipped with DDR4 memory.

On devices with no power throttling, the ideal bandwidth attainable from this HBM device is 512~GB/s.
On such a device, \name{} can operate up to ten decompressors in parallel, only limited by the HBM capacity necessary to store the compression reference genome.
Across 10 decompressors, \name{} would achieve 160~GB/s of total throughput, which is 31\% of the peak HBM bandwidth.
We emphasize again that the compression reference cannot be shared across multiple decompressors, because the primary limitation to an individual decompressor's performance is random access into the compression reference.
Sharing a reference across two decompressors will effectively halve the performance of each decompressor.

We also present the performance comparison against a GPU implementation in the following section, after compression performance evaluation.

\subsubsection{Compression performance}
\begin{figure}[htb]
    \centering
    \includegraphics[width=\columnwidth,page=3]{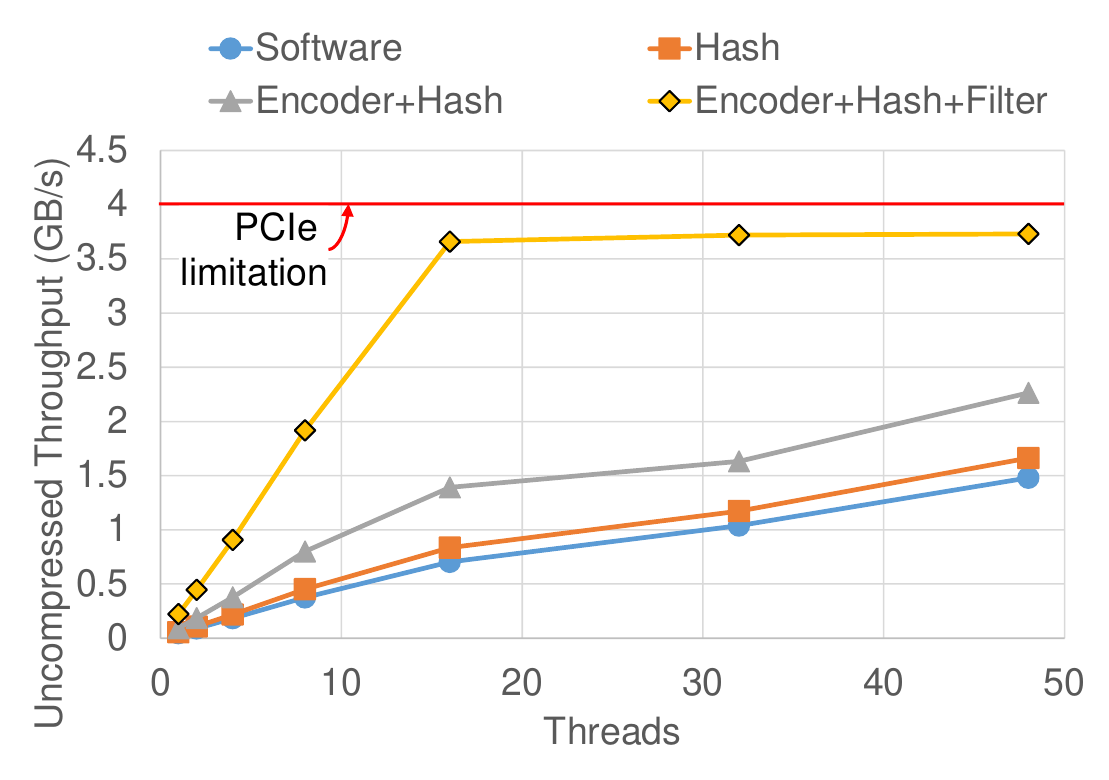}
    \caption{Compression performance with gradual accelerator offloading on the HG002 long read dataset.}
    \label{fig:compression_performance}
\end{figure}

\name{}'s compressor design chose to partition the work between the host software and the accelerator hardware, because the cuckoo hash table used during compression was too large to comfortably keep on the accelerator memory.
Figure~\ref{fig:compression_performance} compares the performance between different partitioning of work, and how many host CPU threads are necessary to reach maximum performance.

We can see that the full software is primarily computation-bound, as performance continuously increases with more threads.
Offloading hashing alone did not have much performance benefits, since the Murmur3 hash we use is actually quite lightweight.
Offloading ASCII parsing (``Encoder'') brings significant performance improvements, but we notice that performance improvements are sublinear after 16 threads.
After encoding and hashing are both offloaded to the accelerator, the only work left on the software is cuckoo hash lookup and matching against the reference.
Reference matching is not the primary performance bottleneck since we can use simple \texttt{memcmp} thanks to the aligned binary representation we presented in Section~\ref{sec:software}.
The reason for sublinear performance in this situation was found to be the 32-bit random access into the cuckoo hash causing excessive cache misses, turning the host memory system into the bottleneck.
\name{} overcomes this using the probabilistic filter, which reduces host CPU and memory pressure and allows the overall system to reach maximum performance limited only by duplex communication over the PCIe, with only 16 CPU threads.

\begin{figure}[htb]
    \centering
    \includegraphics[width=\columnwidth,page=14]{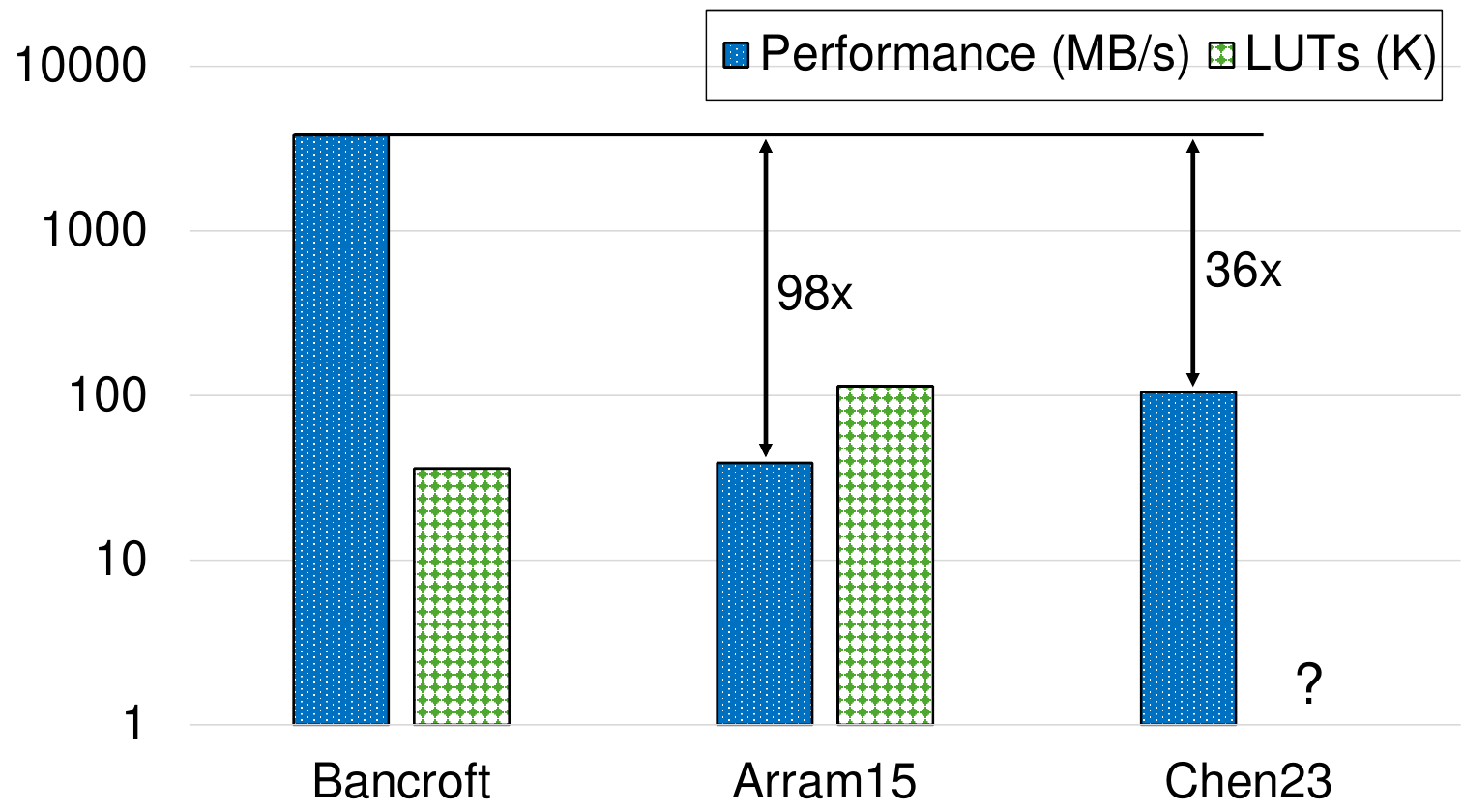}
    \caption{FPGA genome compression comparisons (Log-scale).}
    \label{fig:compressor_comparison}
\end{figure}

The compression performance of \name{} is orders of magnitude higher than any existing genome compression accelerator, as illustrated in Figure~\ref{fig:compressor_comparison}.
We compare against two published genome compression accelerators, Arram15~\cite{arram2015fpgareferencecompressiongenomic} and Chen23~\cite{chen2023efficientsequencingcompressionfpga},
We note that the FPGA platforms used are different and \name{} uses the host memory to host the cuckoo hash, rendering raw LUT count comparison inaccurate.
However, the trend is clear.
\name{} achieves almost 100$\times$ the performance of Arram15, and over 30$\times$ Chen23.
We emphasize that Figure~\ref{fig:compression_efficiency_comparison} is presented in logarithmic scale.

\subsubsection{GPU comparisons}
\begin{figure}[htb]
    \centering
    \includegraphics[width=\columnwidth,page=15]{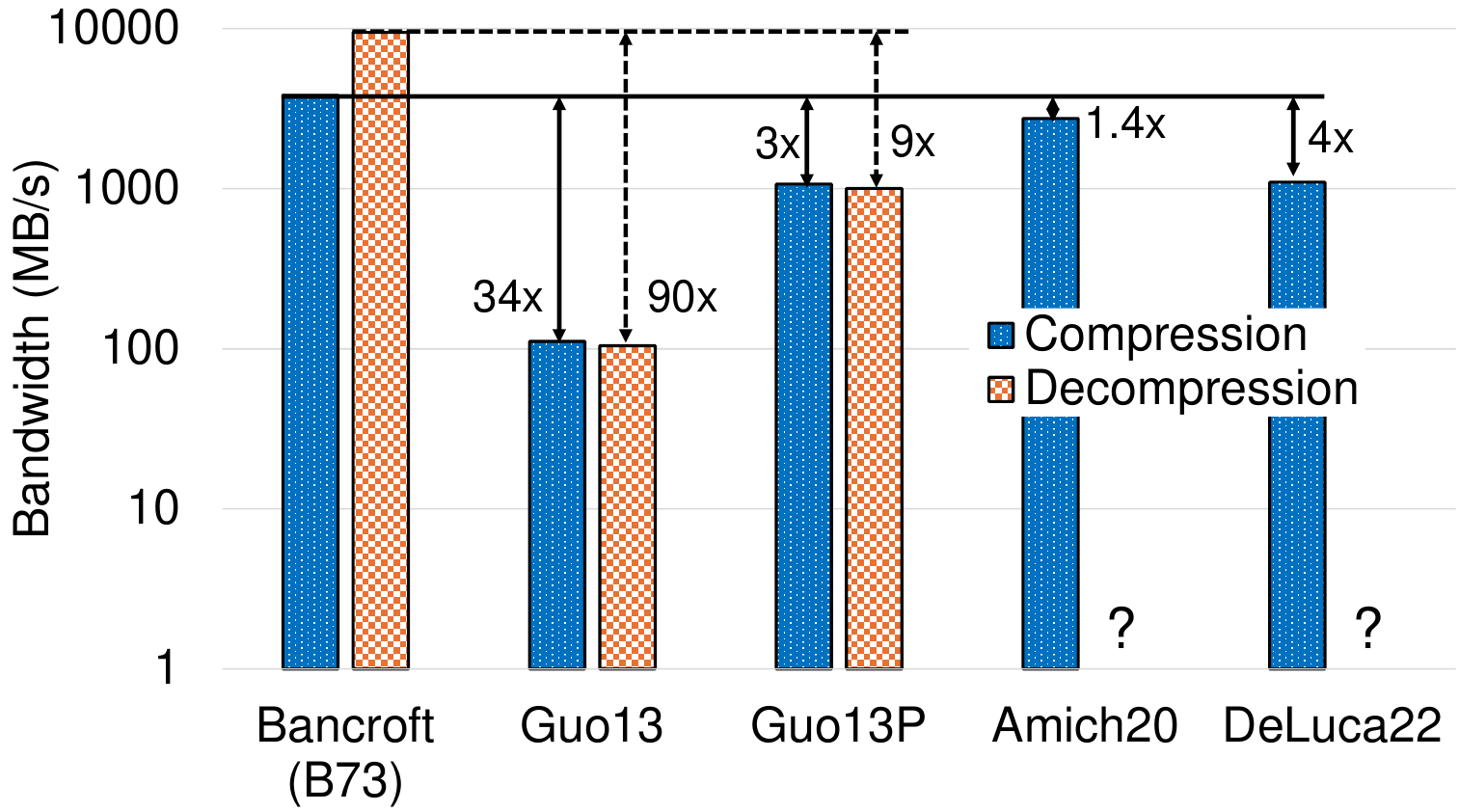}
    \caption{Comparison with GPU-accelerated compression (Log-scale).}
    \label{fig:compressor_comparison_gpu}
\end{figure}

Figure~\ref{fig:compressor_comparison_gpu} compares the compression and decompression performance of \name{} against GPU implementations of genome compression~\cite{guo2013gpu}, based on the worst-case measured performance of single \name{} compression and decompression modules (B73 Maize).
Three systems are compared (Guo13~\cite{guo2013gpu}, Amich20~\cite{amich2020gpu}, DeLuca22~\cite{deluca2022gpu}).
The latter two used multiple current-generation GPUs in parallel, so we present performance normalized to a single GPU.
Guo13 used a now-outdated K20c, so we also present a scaled projection based on the relative performance against a modern RTX~409 GPU (Guo13P).
Only Guo13 presents a decompression performance, which is similar to its compression performance.

\name{} achieves superior performance compared to all systems compared against.
We also emphasize that \name{}'s decompressor consumes only 3\% of the U50 chip, while the GPU implementations use all available GPU resources.

\subsection{Case Study: Pre-Alignment Filtering}

We use pre-alignment filtering acceleration as a driving memory-intensive application.
Sequence alignment is one of the most fundamental components of computational genomics.
A genomic sequence may need to be aligned against one or more references simultaneously or against one or more sequences simultaneously~\cite{leggett2016nanookmultireference,hohl2002efficient,angiuoli2011mugsy,schmidt2019accurate}, for De Novo assembly and evolutionary studies.

For the popular seed-and-extend method of alignment, which is currently the de facto standard, each read can be matched to numerous potentially matching reads in the database, or locations in the reference, during the seeding process.
Most of these matches turn out to be unfruitful after proper alignment.
Pre-alignment filtering helps circumvent the high computational overhead of actual long read alignment by quickly filtering out unlikely matches based on edit distance thresholds, often resulting in an order of magnitude reduction of end-to-end work for both long and short reads, without loss of accuracy~\cite{alser2019shouji,alser2017gatekeeper, alser2017magnet,alser2020sneakysnake}.

We note that, especially for sequence-to-sequence alignment, every new filtering operation needs to transport one or more read sequences to the accelerator.
This is because of the sheer size of the sequence-read databases, as well as the inherently unpredictable nature of sequence-to-sequence matching.
Every potential alignment operation statistically involves a long read which must be newly read and transported to accelerator memory.
For simplicity and consistency, even for sequence-to-sequence alignment, we refer to each of the sequence pairs as ``reference'' and ``read'' from now on.

\subsubsection{Pre-Alignment Filtering Algorithm}
\begin{figure}[htb]
    \centering
    \includegraphics[width=.9\columnwidth,page=10]{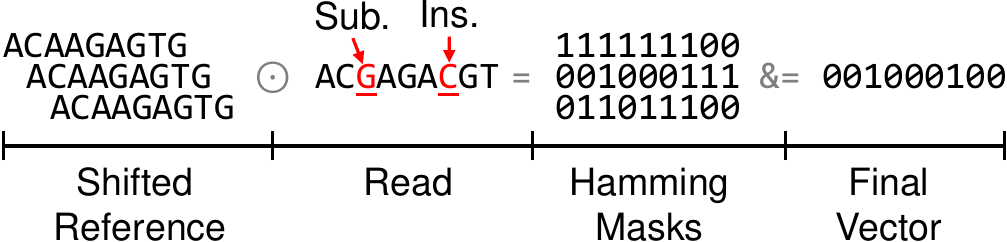}
    \caption{Pre-alignment filtering using Shifted Hamming Distance.}
    \label{fig:shd}
\end{figure}

For evaluation, we implement a Shifted Hamming Distance (SHD) type filter (SHD~\cite{xin2015shiftedhammingdistance}, GateKeeper~\cite{alser2017gatekeeper,bingol2024gatekeepergpu}, MAGNET~\cite{alser2017magnet}), since they have shown to have good and accurate filtering while highly parallelizable, often limited by memory bandwidth~\cite{alser2017gatekeeper}.
Figure~\ref{fig:shd} demonstrates the key idea of SHD filters, filtering based on an edit distance of 1.
A filter with an edit distance of one shifts the reference by one to the left and right, resulting in three shifted copies.
The read sequence (which may have substitutions, insertions, and deletions) is then compared against each copy, to create three-bit masks where ``0'' and ``1'' represent a matched base and a non-match, respectively.
The three masks are \texttt{AND}'d together, and the number of ``1''s remaining becomes the edit distance, which is correctly 2 in this case.
Different designs have introduced numerous optimizations on top of this idea, including turning short runs of zeros (e.g., ``1001'') into ones (``1111'').

We evaluate this application on \name{} using a popular configuration with an edit distance of 5, meaning 11 copies are created by shifting the reference from 1 to 5 times in each direction.
The filtering kernel processes the 11 shifted references and one read all at once using a tree of comparators looking at disjoint windows of 512 bits.
A configurable amendment module after the comparator can flexibly apply various optimizations introduced by algorithms such as MAGNET~\cite{alser2017magnet}.
Then, a tree of saturating 3-bit adders computes the number of 1's in the final vector.

\begin{figure}[htb]
    \centering
    \includegraphics[width=\columnwidth,page=9]{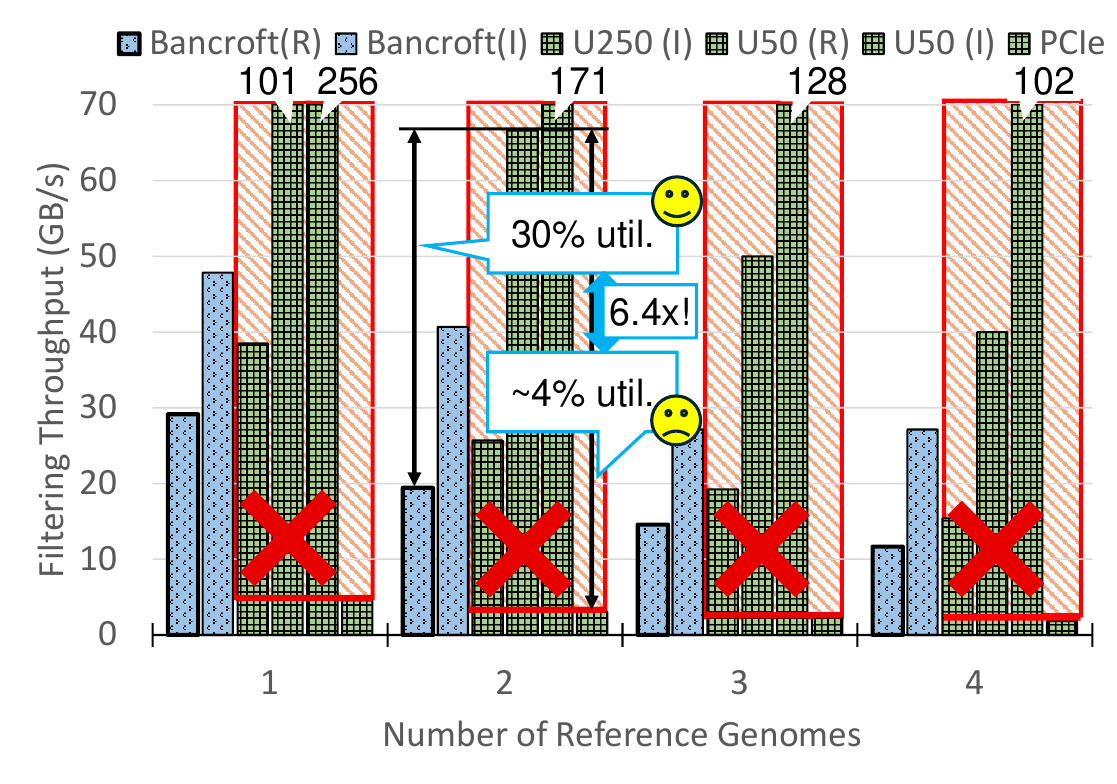}
    \caption{\name{} improves FPGA performance utilization by over 6$\times$ compared to the PCIe bottleneck.}
    \label{fig:filtering_performance}
\end{figure}

\subsubsection{Evaluation Against FPGA Platforms}
Figure~\ref{fig:filtering_performance} presents the performance evaluation of this application on various FPGA accelerator environments.
In addition to simple 1-to-1 mapping, we also consider a multi-reference alignment scenario where a read is mapped to multiple references to improve accuracy or for evolutionary research~\cite{leggett2016nanookmultireference,hohl2002efficient,angiuoli2011mugsy,schmidt2019accurate}.
Compared systems include \name{}(R), the ``real'' implementation of \name{} subject to the U50-specific HBM throttling, and the ``ideal'' \name{}(I), the projected performance without throttling.

This performance is compared against the internal throughput of three different FPGA accelerators, assuming all data already exists in device memory: U250(I), the ideal performance of the Alveo U250 device with DDR4 memory, as well as U50(R) and U50(I), the U50 device where HBM is power-throttled and not, respectively.

But as we repeatedly emphasize in this work, accelerator internal throughput is \textbf{\underline{not representative}} of real-world performance because if the read dataset exceeds memory capacity, every request must transport some data to the accelerator.
In reality, communication overhead over PCIe dominates the runtime for this application.
The hatched red box represents the internal throughput limited by PCIe performance.
The final bar in each group represents the real-world performance considering the PCIe bandwidth bottleneck.

The results show that \name{} achieves over 6$\times$ higher performance compared to the realistic system limited by PCIe performance.
Even compared to the \emph{internal} performance of the in-memory U250(I), \name{} performs delivers over 85\% of the performance.
When compared to the upper-bound performance U50(R) which assumes an unrealistic scenario where where data movement is overhead-free, \name{} delivers performance within the order of magnitude margin, at 30\%.
This means \name{} can operate at 30\% of the peak utilization of the device, while providing practically infinite memory capacity over PCIe.
This is a stark upgrade from the mere 4\% utilization achieved over uncompressed PCIe.

\subsubsection{Evaluation Against GPU Platforms}
We also compare \name{}'s performance numbers against recently published GPU implementations described in Table~\ref{tab:filter_gpus}.
This table lists the publication year, GPU platform used, and the memory bandwidth of the GPU platform.
For reference, HBM on the U50 has a catalog maximum bandwidth of 512~GB/s, but is throttled at 202~GB/s due to power system limitations.
In comparison, all GPU systems have superior memory bandwidth.

\begin{table}[h]
    \centering
    \caption{GPU implementations of pre-alignment filtering.}
    \begin{tabular}{|c|c|c|}
        \hline
        Gatekeeper (GK) & SneakySnake (SS) & SeedHit (SH) \\
        \hline
         2020 \cite{bingol2024gatekeepergpu} & 2024 \cite{alser2020sneakysnake} & 2024 \cite{ju2024seedhitgpu} \\
        GTX 1080ti & RTX 3080ti & RTX 3080ti \\
        484 GB/s & 912 GB/s & 912 GB/s \\   
        \hline
    \end{tabular}
    \label{tab:filter_gpus}
\end{table}

\begin{figure}[htb]
    \centering
    \includegraphics[width=\columnwidth,page=11]{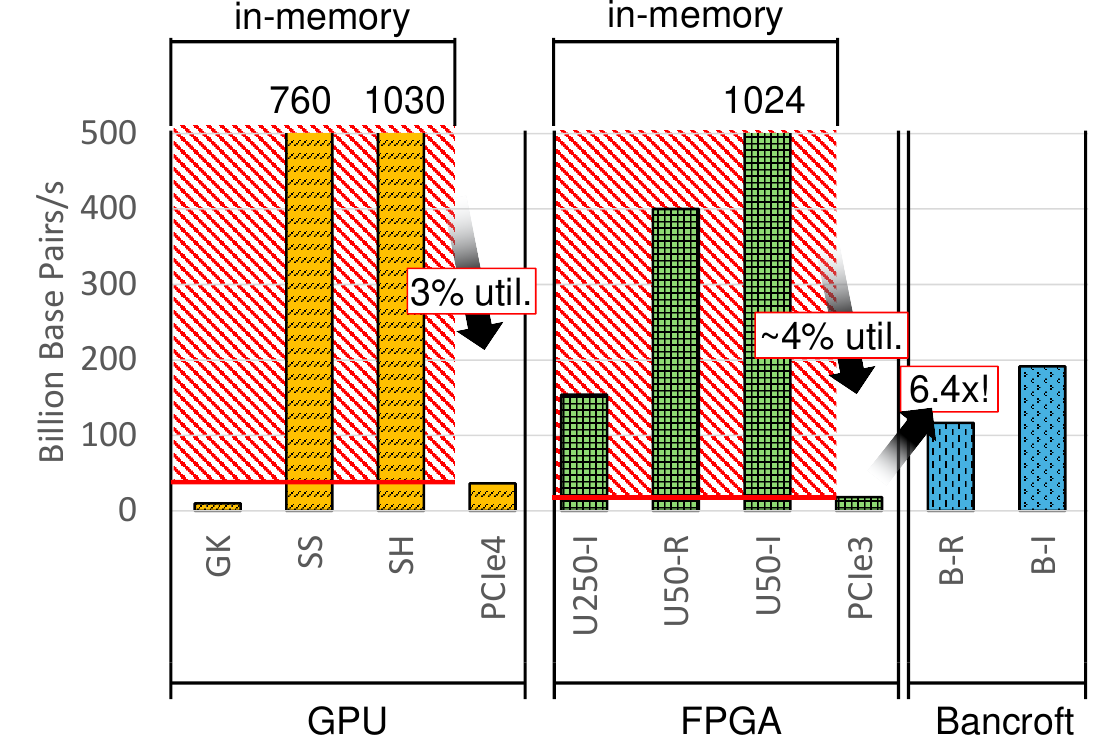}
    \caption{\name{} overcomes PCIe limitations to achieve high performance.}
    \label{fig:gpucomp}
\end{figure}

When GPU platform performance difference between different datasets, we chose the fastest one listed.
The end-to-end performance comparisons can be seen in Figure~\ref{fig:gpucomp}, presented in terms of base pairs processed per second.
\name{}(R) running at 29.2~GB/s on the one-to-one filtering corresponds to 116 billion base pairs compared per second.
We note that Gatekeeper implements an SHD-based algorithm, and the others implement other GPU-optimized algorithms.

The results in Figure~\ref{fig:gpucomp} emphasize the benefits of \name{}.
First, we note that the best internal throughput of both FPGA (U50-I) and GPU (SH) are somewhat similar, despite GPU's higher memory bandwidth.
However, we emphasize again that if the dataset exceeds on-device memory, each request needs to transport either reference, read, or both, over PCIe to GPU memory before analysis.
We note that the GPU systems listed have presented performance with datasets that fit in on-device memory.

Even with the faster PCIe Gen4 $\times$16 available to the RTX~3080ti GPUs, the vast majority of the computational capacity is wasted in real-world scenarios with data transfer requirements, resulting in the very low utilization (3\%) presented in the figure.

\name{} is an attractive solution to overcome this problem, as it significantly outperforms the PCIe-constrained GPU system despite the last-generation PCIe Gen3 limitation.